\documentclass[11pt]{article}

\usepackage{longtable}
\usepackage{makeidx}
\usepackage{fullpage}
\usepackage{amsmath}
\usepackage{amssymb}
\usepackage{setspace}
\usepackage{bbm}
\usepackage{dsfont}
\usepackage{graphics}
\usepackage{epsfig}
\usepackage[font=footnotesize,labelfont=bf,justification=centerlast,width=.94\textwidth]{caption}
 \usepackage{multirow}
\usepackage{array,booktabs}

\usepackage{color}

\usepackage{hyperref}

\hypersetup{
    bookmarks=true,%
    colorlinks,%
    citecolor=blue,%
    filecolor=black,%
    linkcolor=black,%
    urlcolor=blue
}

\newcommand{\hs}{\hspace{0.15mm}}

\newcommand{\be}{\begin{equation}}
\newcommand{\ee}{\end{equation}}
\newcommand{\bes}{\begin{equation*}}
\newcommand{\ees}{\end{equation*}}
\newcommand{\bea}{\begin{eqnarray}}
\newcommand{\eea}{\end{eqnarray}}
\newcommand{\beas}{\begin{eqnarray*}}
\newcommand{\eeas}{\end{eqnarray*}}

\newcommand{\bmat}{\begin{bmatrix}}
\newcommand{\emat}{\end{bmatrix}}
\newcommand{\pd}{\partial}

\def\le{\left}
\def\ri{\right}

\newcommand{\al}{\alpha}

\newcommand{\De}{\Delta}

\newcommand{\mc}{\mathcal}

\newcommand{\VV}{^{(V)}}
\newcommand{\SSS}{^{(S)}}
\newcommand{\mV}{\mathbb{V}}
\newcommand{\mS}{\mathbb{S}}

\def\le{\left}
\def\ri{\right}

\newcommand{\Z}{\mathbb{Z}}

\newcommand{\sA}{\hat {\cal A}}
\newcommand{\sAO}{\hat {\cal A}_0}
\newcommand{\sF}{\hat \Phi}
\newcommand{\sS}{\hat S}

\begin{document}
\numberwithin{equation}{section}
{
\begin{titlepage}
\begin{center}

\hfill \\
\hfill \\
\vskip 0.75in

{\Large \bf The Spectrum of Static Subtracted Geometries}\\

\vskip 0.4in

{\large Tom\'as Andrade${}^a$, Alejandra Castro${}^b$, and Diego Cohen-Maldonado${}^b$}\\

\vskip 0.3in

${}^{a}${\it Rudolf Peierls Center for Theoretical Physics, University of Oxford,\\ 1 Keble Road, Oxford  OX1 3NP, UK} \vskip .5mm
${}^{b}${\it Institute for Theoretical Physics Amsterdam and Delta Institute for Theoretical Physics, University of Amsterdam, Science Park 904, 1098 XH Amsterdam, The Netherlands} \vskip .5mm

\texttt{tomas.andrade@physics.ox.ac.uk, a.castro@uva.nl, d.b.cohenmaldonado@uva.nl}

\end{center}

\vskip 0.35in

\begin{center} {\bf ABSTRACT } \end{center}
Subtracted geometries are black hole solutions of the four dimensional STU model with rather interesting ties to asymptotically flat black holes. A peculiar feature is that the solutions to the Klein-Gordon equation on this subtracted background can be organized according to representations of the conformal group $SO(2,2)$. We test if this behavior persists for the linearized fluctuations of gravitational and matter fields on static, electrically charged backgrounds of this kind. We find that there is a subsector of the modes that do display conformal symmetry, while some modes do not. We also discuss two different effective actions that describe these subtracted geometries and how the spectrum of quasinormal modes is dramatically different depending upon the action used.  

\vfill

\noindent \today

\end{titlepage}
}

\newpage

\tableofcontents

\section{Introduction}

Our understanding of microscopic properties of extremal and supersymmetric black holes are far superior than our understanding of their non-extremal counterparts. The advantage of the extremal solution  is that we can decouple the near horizon geometry \cite{FerraraKalloshStrominger1995,FerraraKallosh1996,FerraraKallosh1996a}, i.e. we can place an extremal black hole in a box. This box not only isolates the horizon, but it as well enhances the symmetries of the geometry suggesting a dual description in terms of a CFT$_2$. This is the core of the Kerr/CFT correspondence \cite{GuicaHartmanSongEtAl2009}, which is a proposal for the microscopic dual of the extreme Kerr solution.

Stretching the proposal of Kerr/CFT a step further, it is tempting to think of the non-extremal black hole as a finite temperature excitation of the CFT describing the extremal solution.  In an attempt to realize this idea,  it was noticed in \cite{Castro:2010fd} that, at low frequencies, linearized fluctuations around the Kerr black hole display a hidden conformal symmetry. More concretely, the solutions to the wave equation organize themselves in representations of the $SO(2,2)$ group in the same fashion as the three dimensional BTZ black hole \cite{MaldacenaStrominger1998a,BalasubramanianKrausLawrence1999}. Despite the fact that the symmetry is only manifest in a low energy limit, it was robust enough to express the Bekenstein-Hawking entropy of the Kerr solution as the statistical entropy of a CFT$_2$ at high temperature \cite{Castro:2010fd,ChenXueZhang2013,ChenLiuZhang2012,CastroLapanMaloneyEtAl2013a}. That is, one could express universally the area law as a Cardy formula, giving support to the Kerr/CFT proposal. 

The drawback of this  proposal  is that the conformal features of the fluctuations is too fragile: as we move away from the low energy regime there is very little evidence that exploiting the $SO(2,2)$ symmetry is the correct way to describe the black hole \cite{CastroLapanMaloneyEtAl2013}. A rather interesting way to overcome this obstacle was proposed in \cite{Cvetic:2011hp,Cvetic:2011dn}. The authors there suggested a concrete way to put a non-extremal black hole in a box. Remarkably, this idea realizes the hidden conformal symmetry in \cite{Castro:2010fd}  for probe scalars without relying on a low frequency limit. The important feature of this box is that it doesn't tamper with the horizon of the original configuration: this suggests that the microscopic model that accounts for the entropy is unchanged after placing the box. 

The solutions in \cite{Cvetic:2011hp,Cvetic:2011dn} are known as \emph{subtracted geometries}: the box is constructed by subtracting certain metric factors from the asymptotically flat black hole  solution.  The subtracted geometry is a solution to ${\cal N}=2$ supergravity, and this allows to build these geometries in a variety of ways. For instance, they can be obtained by using solution generating techniques \cite{Virmani:2012kw,Sahay:2013xda,Cvetic:2013vqi} or scaling limits \cite{Cvetic:2012tr}. It is possible as well to build  interpolating solutions  between the asymptotically flat black hole and the subtracted one \cite{Baggio:2012db}.  Various properties of the subtracted geometries have been analysed in the literature. For instance, the thermodynamical properties of the solutions \cite{Cvetic:2014nta,An:2016fzu} and 
holographic renormalization \cite{An:2016fzu} have been worked out. In addition, the behaviour of minimally coupled scalars on this background has been considered in \cite{Cvetic:2014ina,Cvetic:2014eka}.  See as well \cite{Chakraborty:2012nu,Chakraborty:2012fx} for a discussion on the attractor mechanism for subtracted geometries.

Our goal here is to understand dynamical properties of static subtracted geometries. In particular, we will study linearized fluctuations of the gravitational and matter modes that support the subtracted black hole. Along the way, we will report the scaling dimensions of the fluctuations and their quasinormal frequencies.  The general subtracted geometry can carry angular momentum, electric and magnetic charges, in addition to mass. We will not include angular momentum in the backgrounds considered here: cases that are only electrically charged will have enough  structure to illustrate intricate properties of the fluctuations. Nevertheless, it would be interesting to add rotation and see which features we find here persist.  

These fluctuations will  test if the hidden conformal symmetry persists for perturbations that are not necessarily minimally coupled. Unfortunately, we will see that this symmetry is only present in certain sectors. 
More broadly, a complete understanding of the fluctuations  can  provide useful information about a potential holographic dual. Given that our analysis can be 
performed analytically to a large extent, it would be very interesting to understand properties of the dual theory. One reason to do so is that some features of the subtracted geometries are also present in other holographic setups, such as those
in Schrodinger spacetimes with $z = 2$ \cite{Andrade:2014kba} and hyperscaling violating solutions \cite{Chemissany:2014xsa}. 
  Here we will only present the bulk analysis of the fluctuations and highlight certain features of the fluctuations; we leave a more holographic analysis for future work.
\subsection{Summary of results}

The main portion of our work involves rather technical analysis of linearized fluctuations. Here we will summarize the 
key points of our method and highlights of our results.  We will also comment briefly on future directions. 

We will build the linearized fluctuations around static subtracted geometries; these geometries are described in Section \ref{sec:thy}. 
To construct the master fields and their equations we will use the technique developed by Kodama-Ishibashi \cite{Kodama:2003jz,Kodama:2003kk}; this analysis is done in Section \ref{sec:fluc}. The strength of this method is that it exploits in a clever manner gauge invariance and isometries to get decoupled ODEs for the physical modes. The drawback is that spherical symmetry is crucial: for this reason we will only analyze static solutions that carry only electric charge. The modes will be decomposed in vector and scalar modes with respect to spherical harmonic decomposition. Our emphasis will be on  finding solutions to the master field equations and the QNM frequencies. 

We will work with two different actions that contain a static subtracted geometry as a solution: the STU model and an Einstein-Maxwell-Dilaton (EMD) model. At the level of matter fields, the difference between the two theories is that STU contains an axion field, $\chi$, whereas EMD does not. Throughout our analysis we will compare the results for each theory: even though they are very closely related at the level of the action, the structure of the fluctuations will be rather different.  

The quick summary of our results is: 
\begin{description}
\item[Vector Sector.] Here is where we find the most striking difference between the STU model and EMD. Due to a constraint arising from the equation of motion for the axion field, the STU model has no vector excitations. On the other hand, for EMD this sector is non-trivial and the coupled set of ODEs is given in \eqref{eq:mastervector}. We have solved for the quasinormal frequencies of this system numerically and the results are in Fig. \ref{vector QNM}. The frequencies have both a real and imaginary part. 
\item[Scalar Sector.] For the STU model we can consistently take $\delta \chi=0$; here both the STU and EMD model will give the same results. When $\delta \chi=0$ there are eight non-trivial branches of solutions for the modes. For four of these branches the solutions are hypergeometric functions and hence the quasinormal frequencies are integer spaced (and purely imaginary); see \eqref{A gen soln} and \eqref{w A}. For the other four branches, the modes are Heun functions, but rather surprisingly the quasinormal frequencies are still integer spaced and purely imaginary \eqref{omega A=0}.  For $\delta \chi\neq0$, which only applies for the STU model, the modes are again Heun functions. However the quasinormal modes are not integer spaced; see Fig. \ref{axion QNM}. 
\end{description}

It is important to emphasize that for one scalar subsector the fluctuations are appropriately weighted hypergeometric functions: this indicates that $SO(2,2)$ is the natural symmetry to organize this portion of the spectrum.  Since the geometries in consideration have no obvious conformal isometries, it is highly non-trivial that this is occurring. However, there are modes that deviate significantly from this conformal pattern: the solutions in this case are Heun functions instead of hypergeometric functions. We don't have an alternative holographic interpretation of this sector at the moment; we just know it does not smell like a CFT and it does not mimic the fluctuations of BTZ black hole \cite{BirminghamSachsSolodukhin2002} in the way it does for minimally coupled scalars. It would be interesting to study further what are the basic features of the dual theory based on our results and complement them with the analysis in \cite{Cvetic:2016eiv}.  In particular, some of the non-conformal modes are in the gravitational sector and they would contribute to the energy-density correlation functions. It would be interesting to analysis two-point functions of the stress tensor on this background.

It is possible to uplift a subtracted geometry from four to five dimensions. Rather interestingly, in five dimension the solution is locally AdS$_3\times S^2$ \cite{Cvetic:2011dn}. This suggests that the modes should be organized using the conformal symmetry of AdS$_3$, but we do not find evidence of this from the four dimensional point of view.  However the uplift is done in the magnetic frame, whereas we are always working in the electric frame, and this might obscure certain properties. For instance, there could be a non-trivial arrangement of the couplings as we uplift that restores the conformal features in the five dimensional geometry, but this is highly speculative. In this work we only discuss four dimensional properties of the solution.

\section{The theory and the solution}\label{sec:thy}

In this section we will lay down the main features of the theory we will analyze, and more importantly, the solutions we will focus on. Our conventions mostly follow those in \cite{Baggio:2012db,An:2016fzu}. Our theory will be a truncation of the STU model \cite{Cremmer:1984hj,Duff:1995sm}, for which the matter content involves two gauge fields, a scalar and axion field. The action for this truncation is
\begin{align}\label{eq:elecaction}
\nonumber
	I =& {1\over 16\pi G}\int d ^4 x \sqrt{g} \bigg(  R - \frac{3}{2} \partial_\mu \eta\partial^\mu \eta - \frac{3}{2} e^{2 \eta} \partial_\mu \chi  \partial^\mu \chi- \frac{1}{4} e^{- 3 \eta} (F^0)^2 
	- \frac{3}{4} \frac{e^{- \eta}}{(4 \chi^2 + e^{- 2 \eta})} (\tilde F - \chi^2 F^0)^2 \\
	&  -\frac{\chi}{(4 \chi^2 + e^{- 2 \eta})} \left[ 3 \tilde F \wedge \tilde F + 3(2 \chi^2 + e^{- 2 \eta}) \tilde F \wedge F^0 
	- \chi^2 (\chi^2 + e^{- 2 \eta}) F^0 \wedge F^0 \right]  \bigg)~.
\end{align}
This is known as the \emph{electric frame} action; equations of motion and some conventions are presented in Appendix \ref{app:eom}. The most commonly known version of this model is usually written in the \emph{magnetic frame}, in which it reads
\begin{align}\label{eq:smag}
\nonumber
	I =& {1\over 16\pi G} \int d ^4 x \sqrt{g} \bigg(  R - \frac{3}{2} \partial_\mu \eta\partial^\mu \eta- \frac{3}{2} e^{2 \eta}\partial_\mu \chi  \partial^\mu \chi- \frac{1}{4} e^{- 3 \eta} (F^0)^2 
	- \frac{3}{4} {e^{- \eta}}( F + \chi^2 F^0)^2 \\
	&  + 3 \chi F \wedge  F + 3\chi^2 F \wedge F^0 + \chi^3 F^0 \wedge F^0  \bigg )~.
\end{align}
The relation between both of them is given by 
\be
F_{\mu\nu}=-(4\chi^2+e^{-2\eta})^{-1}\le({1\over 2} \varepsilon_{\mu\nu\rho\sigma}e^{-\eta}(\tilde F-\chi^2 F^0)^{\rho\sigma} +2\chi \, \tilde F_{\mu\nu}+\chi (2\chi^2 +e^{-2\eta})F^0_{\mu\nu}\ri)~.
\ee
Here we will mostly use the electric frame: the reason simply being that the matter content will respect the spherical symmetry of the background solutions we will consider. 

The backgrounds that we will present below will all have $\chi =0$. This is not a consistent truncation of the STU model (in either frame), since  setting $\chi=0$ in the equation of motion  gives a constraint  between the remaining fields:
\begin{equation}\label{constraint}
	\tilde F \wedge \tilde F - e^{- 2 \eta} F^0 \wedge \tilde F = 0~.
\end{equation}
However, it is interesting to note that for configurations with $\chi=0$, the equations of motion for the remaining fields can be obtained from the following action
\begin{equation}\label{eq:ieff}
	I_{\rm eff} =  {1\over 16\pi G}\int \sqrt{g} \left[ R - \frac{3}{2} \partial_\mu \eta\partial^\mu \eta - \frac{e^{- 3 \eta}}{4} (F^0)^2 - \frac{e^{\eta}}{4}\tilde F^2 \right]~.
\end{equation}
We will refer to this theory as EMD: Einstein-Maxwell-Dilaton theory. Throughout our analysis, we will contrast the results obtained from using \eqref{eq:elecaction} versus \eqref{eq:ieff}. 

\subsection{Static subtracted geometries}

The focus of our work is to study dynamical properties of a specific class of solutions to \eqref{eq:elecaction}: electrically charged black holes which are asymptotically conical. These solutions are known as subtracted Reissner-Nordstrom (subRN) geometries, since they were first constructed by subtracting certain metric factors from the asymptotically flat Reissner-Nordstrom  solution \cite{Cvetic:2011hp,Cvetic:2011dn}. More generally, these solutions can be obtained using solution generating techniques \cite{Virmani:2012kw,Sahay:2013xda,Cvetic:2013vqi}, scaling limits \cite{Cvetic:2012tr} or interpolating solutions \cite{Baggio:2012db}. In the following we will summarize some basic properties of subRN background.

The asymptotically flat Reissner-Nordstrom with electric and magnetic sources is given by
\begin{align}\label{eq:RN}
	ds^2& = \frac{\sqrt{\Delta_{\rm RN}}}{X} dr^2 - \frac{X}{\sqrt{\Delta_{\rm RN}}} dt^2 + \sqrt{\Delta_{\rm RN}} d \Omega^2_2~, 
	 \cr
	e^{\eta} &= \sqrt{\frac{p(r)}{p_0(r)}}~, \qquad \chi = 0~, \qquad A^0  = {m \sinh(2\delta_0)\over p_0(r)}dt~ , \qquad A=m \sinh(2\delta)\cos\theta d\phi~,
\end{align}
where 
\begin{align}\label{eq:RNf}
&\Delta_{\rm RN}(r)= p(r)^3 p_0(r)~, \quad X(r)=r^2-2mr~,\cr &p(r)=r+2m \sinh^2\delta ~,\quad p_0(r)=r+2m\sinh^2\delta_0~.
\end{align}
Here $m$, $\delta$ and $\delta_0$ are constants. This solution asymptotes to $\mathbb{R}^{3,1}$ and has an inner and outer horizon located at the zeroes of $X(r)$. The conserved charges of this black hole, i.e. mass, electric and magnetic charge, are
\be
M={m\over 4G}(\cosh (2\delta_0) +3 \cosh (2\delta) ) ~,\quad Q_{\rm elec}={m\over 4G}\sinh(2\delta_0)~, \quad Q_{\rm mag}={3m\over 4G}\sinh(2\delta)~.
\ee
The Hawking temperature reads
\begin{equation}
	T = \frac{1}{8 \pi m}  ( \cosh \delta_0 \cosh^3 \delta  )^{-1}~,
\end{equation}
\noindent and the entropy is given by
\be\label{eq:SBH}
S_{\rm BH}={A_H\over 4G}={4\pi m^2\over G} \cosh\delta_0\cosh^3\delta~.
\ee

A so-called subtracted version of \eqref{eq:RN} is given as follows. In the magnetic frame, the subtracted solution takes the form
\begin{align}\label{eq:sbRNM}
	ds^2 &= \frac{\sqrt{\Delta}}{X} dr^2 - \frac{X}{\sqrt{\Delta}} dt^2 + \sqrt{\Delta} d \Omega^2_2~, \cr
	e^\eta &= \frac{B^2}{\sqrt{\Delta}}~, \qquad \chi = 0~, \qquad A^0  = \frac{2m B^3\Pi_s \Pi_c}{(\Pi_c^2 - \Pi_s^2)} \Delta^{-1} dt~ , \qquad  A = B \cos\theta d\phi ~,
\end{align}
with $X(r)$ as in \eqref{eq:RNf} and
\begin{equation}
 \Delta(r) = (2m)^3(\Pi_c^2 - \Pi_s^2) r + (2 m)^4 \Pi_s^2~.
\end{equation}
The parameters $\Pi_{c,s}$ $m$ and $B$ are constant. The horizons of this solution are again given by the zeroes of $X(r)$. Note that this solution, as $r\to \infty$, takes the form
\begin{align}\label{eq:vacsol}
	ds^2 &=  \sqrt{r}\le(\ell^2 {dr^2\over r^2}-{r\over \ell^2} dt^2 + \ell^2 d\Omega_2^2\ri)~, \qquad \ell^2= \sqrt{ (2m)^3(\Pi_c^2 - \Pi_s^2)}~, \cr
	e^\eta &= \frac{B^2}{\ell^2 \sqrt{r}}~, \qquad \chi = 0~, \qquad	A^0  = 0 ~ , \qquad  A = B \cos\theta d\phi ~,
\end{align}
which we would identify as a ``vacuum solution'' to \eqref{eq:sbRNM} and as such it gives a reference point to quantify observables \cite{An:2016fzu}. This solution has interesting scaling properties, which mimics those in hyperscaling violating geometries \cite{Huijse:2011ef,Dong:2012se}. However, it is a singular solution to the system and hence it is only used as an asymptotic solution.   

As we mentioned above, \eqref{eq:RN} has an intimate relation to \eqref{eq:sbRNM} via solution generating techniques, scalings, or explicit subtractions. This relates the parameters in the solutions as
\be\label{eq:rel}
\Pi_s=\sinh\delta_0 \sinh^3\delta~,\qquad \Pi_c=\cosh\delta_0 \cosh^3\delta~, \qquad B=2m\sinh\delta~.
\ee
This relation in particular assures that the entropy of both black holes is exactly the same and given by \eqref{eq:SBH}. Furthermore, 
the surface gravities of both black holes are as well the same (provided the same Killing vector is used for both solutions).  However, the subRN solution has its own conserved charges: using canonical definitions within the framework of holographic renormalization, the physical quantities associated to \eqref{eq:sbRN} are \cite{An:2016fzu} 
\be
M={(2m)^4\over8G \ell^4}(\Pi_c^2+\Pi_s^2 ) ~,\quad Q_{\rm elec}={(2m)^2\Pi_c\Pi_s\over 4G B^3}~, \quad Q_{\rm mag}={3B\over 4G}~.
\ee

For the purpose of studying the fluctuations around subRN, it is more convenient to have a electrically charged solution. The  subtracted version of \eqref{eq:sbRNM} in the electric frame  is given by
\begin{align}\label{eq:sbRN}
	ds^2 &= \frac{\sqrt{\Delta}}{X} dr^2 - \frac{X}{\sqrt{\Delta}} dt^2 + \sqrt{\Delta} d \Omega^2_2~, \cr
	e^\eta &= \frac{B^2}{\sqrt{\Delta}}~, \qquad \chi = 0~, \qquad A^0  = \frac{2m B^3 \Pi_s \Pi_c}{(\Pi_c^2 - \Pi_s^2)} \Delta^{-1} dt~ , \qquad \tilde A = - \frac{1}{B}(r - 2m) dt ~,
\end{align}
which is a solution to both \eqref{eq:elecaction} and \eqref{eq:ieff}. This will be the solution we will use throughout our analysis in the following section, and we emphasize that we will not use \eqref{eq:rel}: the parameters in \eqref{eq:sbRN} should be thought to be independent. Finally, for sake of simplicity, in Section \ref{sec:fluc} we will set $B=2m$. Shifting the value of $B$ can be acomplished by shifting $\eta$ by a constant and rescaling the field strengths appropriately, i.e. 
\be
\eta~ \to~ \eta + \eta_0 ~,\qquad \tilde F~\to~ e^{-\eta_0/2}\tilde F ~, \qquad  F^0~\to~ e^{3\eta_0/2}F^0~, 
\ee
with $\eta_0$ constant. This is a symmetry of the equations of motion (for $\chi=0$) and hence qualitative aspects of our results are not impacted by making the choice $B=2m$.  

 
When $\Pi_s = 0$, $\Pi_c=1$ and $B=2m$ we obtain a version of subtracted Schwarzschild \cite{Yuan:2013ts}. The solution is conformal to $AdS_2 \times \mathbb{R}^2$ as can be checked explicitly from the above expressions. As we study the fluctuations we will consider this a limit case, and due to the explicit symmetry of the background  all perturbations can be solved for exactly in terms of Hypergeometric functions.
See \cite{Anninos:2011af, Anantua:2012nj, Davison:2014lua} for related examples. 

\section{Linearized fluctuations}\label{sec:fluc}

In this section we study the linearized fluctuations of the metric and matter fields of subRN in the electric frame of the STU model and the EMD theory described section \ref{sec:thy}. The background solution we will always consider is \eqref{eq:sbRN}. 

\subsection{Warm-up: minimally coupled scalars}\label{sec:min}

Before proceeding, it is instructive to review the key property that initially motivated the construction of the subtracted geometries: the dynamics of a probe scalar field. Prior analysis similar to the one below are given in \cite{Cvetic:2011dn,Cvetic:2014ina}. The behavior of this field should be contrasted with the  modes in \eqref{eq:genmodes} which we will derive in the following section. Consider a massless and neutral scalar field; its Klein Gordon equation is
\be\label{eq:KG}
{1\over \sqrt{-g}}\partial_\mu \le(\sqrt{-g} g^{\mu\nu}\partial_\nu \Psi \ri)=0~.
\ee 
Expanding in eigenmodes and using separability
\be
\Psi(x^\mu)= e^{-i \omega t} \mS(\theta, \phi)  R(r) ~,
\ee
gives that \eqref{eq:KG} reduces to
\be\label{eq:hgmin}
R'' + \frac{X'}{X} R' + \left( - \frac{\ell(\ell+1)}{X} + \frac{\omega^2 \Delta}{X^2} \right) R =0  ~. 
\ee
\noindent Here primes denote derivatives with respect to $r$, $\mS$ are the usual spherical harmonics on $S^2$ defined by 
\begin{equation}
	( \hat \nabla^2 +  \ell(\ell+ 1) ) \mS = 0~,
\end{equation}
\noindent with $ \hat \nabla^2$ the Laplacian on $S^2$ and $l\in\Z^+$. Introducing
\begin{equation}\label{redef r w}
	y = {2 m\over  r} ~ , \qquad \hat \omega = {\omega \over 4 \pi T} ~,
\end{equation}
\noindent where the Hawking temperature for subRN is given by $T = (8 \pi m \Pi_c)^{-1}$, we verify that \eqref{eq:hgmin} depends on the charges only via 
\begin{equation}
 	\epsilon \equiv \frac{\Pi_s}{\Pi_c}~,
 \end{equation} 
\noindent and takes the form 
\begin{equation}\label{KGeq2}
 (y-1)R'' + R' + \left( \frac{\ell(\ell+1)}{y^2} +  \frac{\hat \omega ^2 \left((y-1) \epsilon ^2+1\right)}{(y-1) y} \right) R = 0~,
\end{equation}
\noindent where primes now denote derivatives with respect to $y$. 
The solution to \eqref{KGeq2} can be written in terms of hypergeometric functions as
\begin{align}
	R = & C_1 y^{\ell+1} (1 - y)^{- i \hat \omega} \hs _2 F_1(  \ell + 1 - i \hat \omega (1 + \epsilon) ,   \ell + 1 - i \hat \omega (1 - \epsilon) , - 2 \ell , y )  \cr
	&+C_2 y^{-\ell} (1 - y)^{- i \hat \omega} \hs _2 F_1( - \ell - i \hat \omega (1 + \epsilon) ,  - \ell - i \hat \omega (1 - \epsilon) , - 2 \ell , y )~. 	
\end{align}
Note that the scaling dimensions of the scalar, i.e. the characteristic exponents of the power law as $y\to 0$, depend on the quantum 
numbers of the fields. This is very common occurrence in the near horizon geometries of extremal black holes, and more generally in cases where the metric is a direct product. It is as well present in geometries with non-trivial scaling properties in the UV, such as the cases studied in, for example,  \cite{GuicaSkenderisTaylorEtAl2011,CompereGuicaRodriguez2014,Chemissany:2014xsa}. This feature is usually interpreted as some semi-local behavior of the dual theory \cite{Iqbal:2011in}.  
We will encounter a similar type of dependence for the scaling dimensions of all metric and matter fluctuations, and it would be interesting to account for this holographically.

The quasi-normal modes (QNMs) of the black hole under consideration are solutions to the linearized equations of motion which satisfy regularity at the boundary $y=0$ and ingoing boundary conditions at the horizon $y=1$ \cite{Horowitz:1999jd, Berti:2009kk}.
The latter condition corresponds to a near horizon behaviour of the form $R \sim (1 - y)^{- i \hat \omega}$ times a regular power series in $y$. 
These requirements imply that the QNM frequencies are given by 
\begin{equation}
	\hat \omega = - \frac{i}{1 \pm \epsilon} (\ell + n)~, \qquad n = 0, 1, 2, \ldots~.
\end{equation}

From the structure of the background metric \eqref{eq:sbRN}, it is very surprising that the Klein-Gordon equations for a massless field has such a simple solution. One of our goals is to investigate if the coupled metric and matter fluctuations have a similar behavior, and hence which lessons can we draw from a potential holographic dual.

\subsection{Master equations for gravitational fluctuations}
Our starting point is to decompose the fields as
\begin{align}\label{eq:genmodes}
g_{\mu\nu} = \bar g_{\mu\nu} + \delta g_{\mu\nu} ~, \quad \tilde A_{\mu} = \bar {\tilde A}_{\mu} + \delta \tilde A_{\mu} ~,\quad  A_{\mu}^0 = \bar A^0_{\mu} + \delta  A^0_{\mu} ~,  \quad 
\eta = \bar \eta + \delta \eta ~,\quad \chi=\bar \chi +\delta \chi~,
\end{align}
where the barred variables correspond to the background values in \eqref{eq:sbRN} and the pieces proportional to $\delta$ are the fluctuations. The dynamics of these modes are, as expected from the couplings in either \eqref{eq:elecaction} or \eqref{eq:ieff}, a non-trivial coupled system of ODEs. To attack this hurdle we will build master equations by following the techniques in  \cite{Kodama:2003jz,Kodama:2003kk}. In a nutshell, this approach gives a elegant and pragmatic approach to build master field equations for gauge invariant variables by exploiting the  spherical symmetries of the systems. 

Following  \cite{Kodama:2003kk}, we will decompose  further our fluctuations into scalar and vector modes of $S^2$, i.e.
\be
\delta g_{\mu \nu }= h_{\mu\nu}^{(V)} +h_{\mu\nu}^{(S)} ~,\qquad \delta \tilde A_\mu= \tilde A_{\mu}^{(V)}+\tilde A_{\mu}^{(S)}~,\qquad \delta A^0_\mu= A_{\mu}^{0(V)}+A_{\mu}^{0(S)}~,
\ee
 which can be discussed separately. The fluctuations of the dilaton $\eta$ and axion  $\chi$ are all in the scalar sector. Note that there 
 are no tensor perturbations, we cannot build such structures on $S^2$.
 We will as well choose a radial gauge for which 
 \be\label{eq:radial}
 \delta g_{\mu r}=0~,\qquad \delta \tilde A_r=0~, \qquad \delta A_r^0=0~. 
 \ee
This condition does not fully fix the gauge. As we build the master equations, we will build combinations that are invariant under residual diffeomorphisms and  $U(1)$ gauge transformations. 

\subsubsection{Vector modes}

Vector modes are perturbations of the form
\begin{equation}
	\delta g_{ab} = 0~, \qquad \delta g_{a i} = e^{- i \omega t} f\VV_a (r) \mV_i~, \qquad \delta g_{ij}  =  e^{- i \omega t} H\VV (r) \mV_{ij}~, 
\end{equation}
\begin{equation}\label{vec def}
	\delta \tilde A_a = 0 ~, \qquad \delta \tilde A_i = e^{- i \omega t} a (r) \mV_i   ~, \qquad \delta A^0_a = 0 ~, \qquad \delta A^0_i = e^{- i \omega t} b (r) \mV_i ~,
\end{equation}
where $(a,b)=(t,r)$ and $(i,j)=(\theta,\phi)$ and we decomposed the fluctuations in frequency eigenmodes; note that there are  no fluctuations for the dilaton and axion in this sector. Here $\mV_i$ are vector harmonics on $S^2$, which can be simply taken to be 
$\mV_i = \epsilon^{ij} \hat D_j \mS$ with $\mS$ being the standard spherical harmonics and $\epsilon^{ij}$, $\hat D_j$ 
are the Levi-Civita tensor and the covariant derivative on the 2-sphere. Note that they satisfy 
\begin{equation}
	( \hat \nabla^2 + k_V^2) \mV_i  =  0 ~, \qquad \hat D^i \mV_i = 0~,
\end{equation}
where $k_V^2 = \ell (\ell+1) - 1$, with $\ell$ an integer greater or equal to $1$. As explained in \cite{Kodama:2003jz,Kodama:2003kk}, modes with $\ell = 1$ correspond to pure diffeo modes, so we consider $\ell \geq 2$ only.
The $\mV_{ij}$ are define via the symmetrized derivative
\begin{equation}
	\mV_{ij} = - \frac{1}{k_V} ( \hat D_i \mV_j + \hat D_j \mV_i  )~.
\end{equation}
Choosing the standard coordinates on the sphere $d \Omega^2 = d \theta^2 + \sin^2 \theta d \phi^2$, these are given by 
%
\begin{align}
	\mV_\theta& = \csc \theta \partial_\phi \mS~ , \qquad \mV_\phi = - \sin \theta \partial_\theta \mS~, \cr
	\mV_{\theta \theta} &= \frac{\csc \theta}{k_V} ( \cot \theta  - \partial_\theta  ) \partial_\phi \mS~, \cr
	\mV_{\theta \phi} &= - \frac{1}{2 k_V} ( \csc \theta \partial_\phi^2 + \cos \theta \partial_\theta - \sin \theta \partial_\theta^2
	  )  \mS~, \cr
	\mV_{\phi \phi} &= - \frac{1}{k_V} ( \cos \theta  - \sin \theta \partial _\theta ) \partial_\phi \mS~.
\end{align}
The diffeomorphisms that preserve the form of the ansatz in the vector sector are generated by the vector field
\begin{equation}
	\xi_V = e^{- i \omega t} \sqrt{\Delta} \mV^i \partial_i ~.
\end{equation}
This generates the pure gauge mode
\begin{equation}
	f_t = - i \omega \sqrt{\Delta}~, \qquad H\VV = - 2 k_V \sqrt{\Delta}~, \qquad a(r) = b(r) = 0~.
\end{equation}
Moreover, it is clear that $a(r)$ and $b(r)$ are invariant under the $U(1)$ gauge transformations associated to the gauge fields. 
Based on this, and closely following \cite{Kodama:2003jz,Kodama:2003kk}, the gauge invariant combinations 
of the fluctuations we will use are $a(r)$, $b(r)$ in \eqref{vec def}, 
in addition to $\mc W$ defined by
\begin{align}
\mc W(r)\equiv \frac{X}{2 i k_V \omega \Delta^{1/2}} \left( (H^{(V)})'  -   \frac{\Delta'}{2}  H^{(V)}   \right)~,
\end{align}
where prime denotes derivative with respect to $r$. The remaining component of the metric perturbation, $f_t$, can be written in terms 
of $\mc W$ using the equations of motion.
At the linearized level, the Einstein equations and the Maxwell equations for both gauge fields gives the following system of coupled equations
\begin{align}\label{eq:mastervector}
&
\mc W'' +\left(\frac{X'}{X}-\frac{\Delta '}{\Delta}\right)\mathcal{W}' +\left(\frac{1-k_V^2}{X}+\frac{\omega ^2 \Delta }{X^2}\right)\mc W
+\frac{6 m}{X} a
+\frac{ 2m\Pi _c \Pi _s }{X}  b =0~,\cr
&a '' + \left(\frac{X'}{X}-\frac{\Delta '}{\Delta }\right)a '+ \left(-{\left(k_V^2+4\right) \over X}+{\omega ^2 \Delta\over X^2} \right) a 
-\frac{\left(k_V^2-1\right)}{2 m  X} \mathcal{W}
-\frac{\Pi _c  \Pi _s }{X} b =0~,\cr
& b'' +\left(\frac{\Delta '}{\Delta }+\frac{X'}{X}\right) b' +\left(\frac{\omega ^2 \Delta }{X^2}-\frac1{X}\left(1+k_V^2+\frac{(2m)^5\Pi _c^2\Pi _s^2}{\Delta ^2}\right)\right)  b
\cr &-\frac{3  (2m)^5 \Pi _c \Pi _s}{X\Delta ^2} a
+\frac{ (2m)^7 \Pi _c  \Pi _s \left(1-k_V^2\right) }{X\Delta^2} \mathcal{W}
=0~.
\end{align}
These equations are valid for both the STU model in \eqref{eq:elecaction}, and the effective action in \eqref{eq:ieff}. However, in the STU model we need as well to take into account the constraint \eqref{constraint}, which comes from the equation of motion of the axion field. This constraint gives
\be
\bar {\tilde A}_t' \, b  + (\bar{ A}^0_t \, ' - 2 e^{2 \bar \eta} \bar{\tilde A}_t')\, a = 0  ~.
\ee
Solving for this constraint and replacing in \eqref{eq:mastervector}, it simple to see that the only possible solution is
\be
 a (r) =b(r)=\mc W(r)=0 ~.
\ee
Hence, {\it all} vector fluctuations in the STU model are trivial. 

However, if subRN is viewed as a background solution to \eqref{eq:ieff}, we don't have additional constraints and the task ahead is to solve \eqref{eq:mastervector}. Performing the redefinitions \eqref{redef r w} in addition to 
\begin{equation}
	\hat a = \frac{1}{2 m} a(r)~ , \qquad \hat b = \frac{1}{2 m \Pi_c^2} b(r)~,
\end{equation}
\noindent the vector equations \eqref{eq:mastervector} read
\begin{align}
\nonumber
	&\mc W'' + \left( \frac{\left(y^2 \epsilon ^2-2 y \left(\epsilon ^2-1\right)+\epsilon ^2-1\right) }{(y-1) y \left((y-1) \epsilon ^2+1\right)}
	 \right)\mc W' + \left(  \frac{k_V^2-1}{(y-1) y^2}+\frac{\hat \omega^2 \left((y-1) \epsilon ^2+1\right)}{(y-1)^2 y}  \right) \mc W 
	 \\
	 \label{vec1} 
	 &+ \frac{\epsilon  }{y^2(1-y)} \hat b + \frac{3}{y^2(1-y)} \hat a = 0 ~,\\
\nonumber
 &\hat a'' + 
 \left( \frac{\left(y^2 \epsilon ^2-2 y \left(\epsilon ^2-1\right)+\epsilon ^2-1\right) }{(y-1) y \left((y-1) \epsilon ^2+1\right)}  \right) \hat a' 
 + \left( \frac{k_V^2+4}{(y-1) y^2}+\frac{\hat \omega^2 \left((y-1) \epsilon ^2+1\right)}{(y-1)^2 y} \right) \hat a \\
 \label{vec2}
 &+ \frac{\left(1-k_V^2\right)}{(y-1) y^2} \mc W + \frac{\epsilon }{(y-1) y^2} \hat b = 0~, 
 \end{align}
 \begin{align}
\nonumber
&\hat b'' + \left( \frac{\left(\left(y^2-1\right) \epsilon ^2+1\right) }{(y-1) y \left((y-1) \epsilon ^2+1\right)} \right) \hat b' + 
\frac{3 \epsilon }{(y-1) \left((y-1) \epsilon ^2+1\right)^2} \hat a + \left( \frac{\epsilon(1 -k_V^2) }{(y-1) \left((y-1) \epsilon ^2+1\right)^2} \right) \mc W
\\
\label{vec3}
&+ \left( \frac{k_V^2 \left((y-1) \epsilon ^2+1\right)^2+\left(y^2+2 y-2\right) \epsilon ^2+(y-1)^2 \epsilon ^4+1}{(y-1) y^2 \left((y-1) \epsilon ^2+1\right)^2}+\frac{\hat \omega^2 \left((y-1) \epsilon ^2+1\right)}{(y-1)^2 y} \right) = 0~.
\end{align}
The near boundary analysis of \eqref{vec1}-\eqref{vec3} reveals that the characteristic behaviours near $y=0$ are of the form 
$y^{\Delta_V}$ with 
\begin{equation}\label{DV}
	\Delta_V = \left \{ \pm \left( \frac{3}{2} +  k_V^2 \pm \frac{1}{2} \sqrt{13 + 12 k_V^2}  \right)^{1/2} , 1 \pm \sqrt{2 + k_V^2}  
	\right \}~.
\end{equation}

For $\epsilon = 0$, it is possible to decouple \eqref{vec1}-\eqref{vec3} and the resulting equations 
can be solved analytically in terms of hypergeometric functions.
Imposing regularity at the boundary and ingoing boundary conditions at the horizon, we find that the spectrum is given by 
\begin{equation}\label{wV}
	\hat \omega =  -i (\Delta_V^+ + n )~, \qquad n = 0,1,2, \ldots~,
\end{equation}
\noindent where $\Delta_V^+$ are the three positive scaling dimensions in \eqref{DV}.

\begin{figure}[h]
\begin{center}
\includegraphics[scale=0.5]{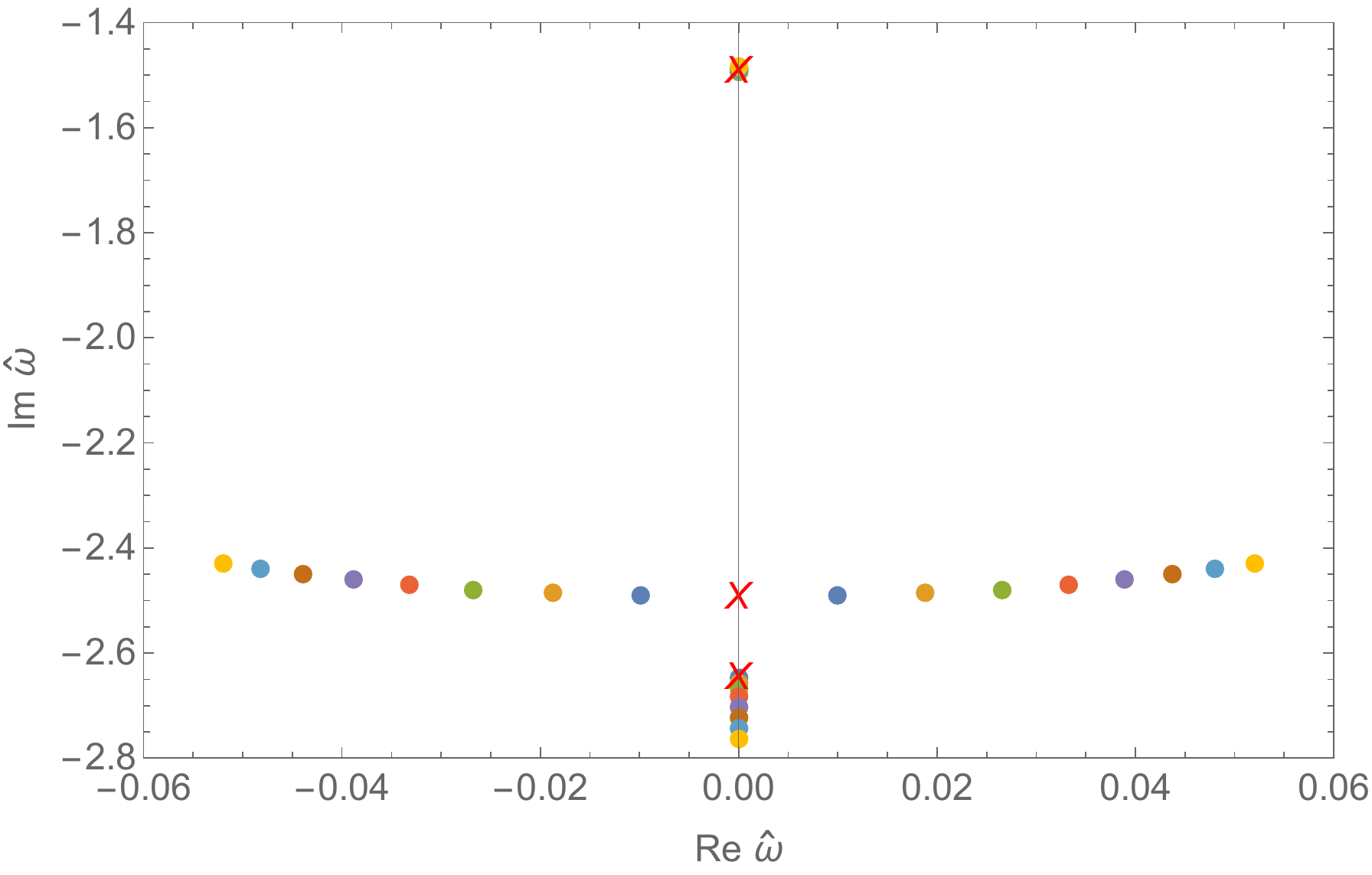}
\caption{Snapshots of the lowest vector QNM in the complex plane for $\epsilon = 0.01,\ldots,0.08$. Different colors correspond to different 
values of $\epsilon$. The red crosses correspond to the analytical values in \eqref{wV}, $\hat \omega =\{1.49265,2.49265,2.64575 \}$.
The first and third QNM remain on the imaginary axis, moving away from each other. The second QNM moves onto the complex plane.}
\label{vector QNM}
\end{center}
\end{figure}

For $\epsilon \neq 0$, it is not clear how to further decouple \eqref{vec1}-\eqref{vec3} while keeping the system of second order. 
Nevertheless, it is rather straightforward to solve the system numerically, and in particular to find its QNMs. 
We do so by discretizing the system of equations and solving the resulting matrix eigenvalue problem numerically. We present 
our results in Fig. \ref{vector QNM}. Note that some of the frequencies acquire a non-zero real part as we increase $\epsilon$, 
manifestly departing from the structure found in the Klein-Gordon equation in section \ref{sec:min}. If the system has a hidden conformal symmetry,  the quasinormal frequencies here should be compared with those in for BTZ black holes \cite{BirminghamSachsSolodukhin2002}: there the frequencies are always interger spaced and purely imaginary. This is not the feature we find here.  

This sharp discrepancy between the two actions in considerations is very interesting: the STU model still supports the conjecture that there is a hidden conformal symmetry in the subRN solution, whereas a different effective action, such as \eqref{eq:ieff}, shows that the quasinormal mode spectrum of subRN in the vector sector does not fit with a conformal description.

\subsubsection{Scalar modes}

We now move on to studying the scalar modes, which are much more intricate. In this case we have for the metric fluctuations
\begin{equation}
	\delta g_{ab} = e^{- i \omega t} h_{ab} \mS~, \qquad 
	\delta g_{a i} = e^{- i \omega t} f\SSS_a (r) \mS_i~, \qquad 
	\delta g_{ij}  =  e^{- i \omega t} ( H\SSS_T (r) \mS_{ij} + H\SSS_L(r) \sigma_{ij} \mS  )~,
\end{equation}
for the gauge fields we take
\begin{equation}
	\delta \tilde A_a = e^{- i \omega t} a_a(r) \mS~, \quad 
	\delta \tilde A_i = e^{- i \omega t} a_\theta (r) \mS_i~, \quad 
	\delta A^0_a = e^{- i \omega t} b_a(r) \mS ~, \quad 
	\delta A^0_i = e^{- i \omega t} b_\theta (r) \mS_i~, 
\end{equation}
and for the dilaton and axion we have
\begin{equation}
	\delta e^{\eta} = e^{- i \omega t} s_1(r) \mS~, \qquad \delta \chi = e^{- i \omega t} s_2(r) \mS~.
\end{equation}
As before, we have $(a,b)=(t,r)$ and $(i,j)=(\theta,\phi)$ and we will use the radial gauge \eqref{eq:radial}. 
Here, $\mS$ are vector harmonics on $S^2$, satisfying
\begin{equation}\label{eq S2}
	( \hat \nabla^2 + k_S^2 ) \mS = 0~,
\end{equation}
\noindent  with $k_S^2 = \ell(\ell+ 1)$ and 
\begin{equation}
	\mS_i = - \frac{1}{k_S} \hat D_i \mS~, \qquad 
	\mS_{ij} = \frac{1}{k_S^2} \hat D_i \hat D_j \mS  + \frac{1}{2} \sigma_{ij} \mS ~.
\end{equation}
Here $\sigma_{ij}$ is the metric on $S^2$.
The eigenvalue equation \eqref{eq S2} implies that $\ell$ is a non-negative integer.  However, modes with $\ell =0, 1$,
are trivial \cite{Kodama:2003jz,Kodama:2003kk} so we focus on  $\ell \geq 2$.
The diffeomorphism that preserves the form of the ansatz in the scalar sector can be written as
\begin{equation}
	\xi = e^{- i \omega t} \Delta^{-1/2} ( c_{t 0}  X  \partial_t + c_{S 0} \Delta \mS^i \partial_i  )~,
\end{equation}
\noindent where $c_{t 0}$ and $c_{S 0}$ are arbitrary constants. This generates the diffeomorphic mode
\begin{align}
	h_{tt} &= - 2 c_{t 0} i \omega X \Delta^{- 1/2} ~, \quad &f^{(S)}_t &= - c_{t 0} k X \Delta^{- 1/2} - c_{S0} i \omega  \Delta^{1/2}~ , \\ 
	\quad H^{(S)}_T &= - 2  c_{S0} k \Delta^{1/2}~, \quad &H^{(S)}_L &= c_{S0} k \Delta^{1/2}~,
\end{align}
\begin{equation}
	a_t =  c_{t0} i \omega \bar {\tilde A}_t~, \quad a_\theta = c_{t0} k \bar {\tilde A}_t~, \quad  
	b_t =  c_{t0} i \omega  \bar A^0_t~, \quad b_\theta = c_{t0} k \bar A^0_t~, 
\end{equation}
\begin{equation}
	s_1 = 0 ~, \quad s_2 = 0~,
\end{equation}
In addition, we record the linearized field strengths since they are invariant under the $U(1)$ gauge transformations
\begin{align}\label{d F 1}
	\delta \tilde F_{r t}	&=  e^{- i \omega t} a_t' \mS , \quad &\delta \tilde F_{r i}&=   e^{- i \omega t} a_\theta' \mS_i~ , 
	\quad &\delta \tilde F_{t i}& = e^{- i \omega t}  (k a_t - i \omega a_\theta ) \mS_i~,  \\ 
\label{d F 2}
	\delta  F^0_{r t}	&=  e^{- i \omega t} b_t' \mS~ , \quad &\delta F^0_{r i} &=   e^{- i \omega t} b_\theta' \mS_i ~, 
	\quad &\delta F^0_{t i}& = e^{- i \omega t}  (k b_t - i \omega b_\theta ) \mS_i ~.
\end{align}
It should be noted, however, that expressions \eqref{d F 1} and \eqref{d F 2} are not invariant under diffeomorphisms. We shall take this
into account when constructing our gauge invariant variables. 

As we saw in the vector modes, the equation of motion for the axion field played a crucial role: it is a constraint that forced the dynamics of all fluctuations to be trivial. However, for the scalar modes its role is a bit different. This equation gives a quadratic equation for $\delta \chi$ which reads 
\be\label{eq:axion}
s_2 '' + \left( {\Delta'\over \Delta} + \frac{X'}{X}  \right) s_2 ' + \left( \frac{\omega^2 \Delta}{X^2} - \frac{k^2}{X} + 
	{2(2m)^4\Pi_s\Pi_c\over \Delta} + 4 \right) s_2  = 0~.
\ee
Note that there is no source from other fluctuations in the scalar modes, the equation for $\delta \chi$ nicely decouples. This gives us two possible routes: we can either set $\delta \chi=0$ and solve for the remaining modes or consider non-trivial solutions to \eqref{eq:axion}. In the following we will consider both cases.

\subsubsection*{Scalar modes with $\delta \chi=0$}
  
Let us study first the scalar modes with $\delta \chi=0$. We again closely follow \cite{Kodama:2003jz,Kodama:2003kk} and write the fluctuation equations in terms of gauge invariant quantities. 
We obtain four gauge invariant variables, each representing the degrees of freedom of the metric, the scalar and each of the gauge fields. 
We denote the fields as $\Phi$ for the metric, $S$, for the scalar, and ${\cal A}$, ${\cal A}_0$ for the gauge fields.
%
%
Their expressions in terms of the basic fields are 
\bea \label{MF1}
\Phi&=&-\frac{iX}{k_S\omega\sqrt\Delta}f_t'^{(S)}+
\frac{4\sqrt\De}{\De'}H_L^{(S)}+
\left(\frac{X'}{2k_S^2\sqrt\De}+2\frac{\sqrt\De}{\De'}-\frac{X\De'}{2k_S^2\De^{3/2}}
\right)H_T^{(S)}\\
&+&\left(
\frac{iX'}{k_S\omega\sqrt\De}-\frac{iX\De'}{2k_S\omega\De^{3/2}}
\right)~,\nonumber\\
\nonumber\\
\label{MF2}
\mc A&=&-\frac{s_1}{2k_Sm}-\frac{H_L^{(S)}}{2k_Sm\sqrt\De}-\frac{H_T^{(S)}}{4k_Sm\sqrt\De}-
\frac{h_{tt}\sqrt\De}{4k_SmX}+\frac{a_t'}{k_S}~, \\
\label{MF3}
\nonumber
\mc A_0&=&\frac{\De^2}{k_S}b_t'+\frac{48m^4\Pi_c\Pi_s\De'}{k_S(\Pi_c^2-\Pi_s^2)}s_1
-\frac{16m^4\Pi_c\Pi_s\De'}{k_S(\Pi_c^2-\Pi_s^2)\sqrt\De}H_L^{(S)}-\frac{8m^4\Pi_c\Pi_s\sqrt\De\,\De'}{k_S(\Pi_c^2-\Pi_s^2)X}h_{tt},\\
&+&\al_1(r)f_t+\al_2(r)H_T^{(S)}+\al_3(r)H_T'^{(S)}+\al_4(r)H_T''^{(S)}~,
\eea
\bea 
\nonumber\\
\label{MF4}
S&=&\frac32k_S\sqrt X\De'\,s_1-\frac{3X^{3/2}\De'^2}{8k_S\De^{3/2}}H_T'^{(S)}+
\frac{3X^{3/2}\De'^3}{16k_S\De^{5/2}}H_T^{(S)},
\eea
where the expressions for the coefficients $\al_i(r)$ can be found in Appendix \ref{app:coef}.

As above we redefine the fields, the radial variable and the frequency such that the equations of motion only
depend on $\epsilon = \Pi_s/\Pi_c$ and $\hat \omega = \omega/(4 \pi T)$, and the radial coordinate $y = 2m/r$. 
This can be achieved by the redefinitions \eqref{redef r w} alongside with 
\begin{equation}
	\sA := {\cal A}~, \qquad \sAO := (2 m)^8 \Pi_c^2 {\cal A}_0~, \qquad \sS := (2 m)^5 \Pi_c^2 S~, \qquad \sF := \frac{(2 m)^2}{k} \Phi~.
\end{equation}
The equations of motion then read 
\begin{align}
\label{scalar eqs 1}
	\sA'' + c_{1  \sA' } (y) \sA' + c_{1 {\cal A} }(y) \sA + c_{1 \sF}(y) \sF+ c_{1 \sF'}(y)\sF' +  c_{1 \sS}(y) \sS  +   c_{1  \sAO }(y) \sAO &= 0~, \\
\label{scalar eqs 2}
	\sAO'' + c_{2 \sAO' } (y) \sAO' + c_{2 \sAO } (y) \sAO + c_{2 \sF}(y) \sF+ c_{2 \sF'}(y) \sF' +  c_{2 \sS}(y) \sS  +   c_{2  \sA }(y) \sA &=0~, \\
\label{scalar eqs 3}
	\sF'' + c_{3 \sF' } (y) \sF' + c_{3 \sF} (y) \sF + c_{3 \sS}(y) \sS + c_{3 \sA}(y) \sA + c_{3 \sAO}(y) \sAO &=0~, \\
\label{scalar eqs 4}
\nonumber
	\sS'' + c_{4 \sS' } (y) \sS' + c_{4 \sS} (y) \sS + c_{4 \sF}(y) \sF+ c_{4 \sF'}(y) \sF' + c_{4 \sA}(y) \sA+ c_{4 \sA'}(y) \sA'  \\ 
	+  c_{4 \sAO'}(y) \sAO' +  c_{4 \sAO}(y) \sAO    &=0 
\end{align}
The coefficients are complicated functions of $y$, $\ell$, $\epsilon$ and $\hat \omega$, and are given in Appendix \ref{app:coef}. 
It is worth mentioning that the frequency dependence occurs only through $\hat \omega^2$, so the equations of motion we 
have obtained can be thought to be of second order in time, as in \cite{Kodama:2003jz,Kodama:2003kk}. Note that the scalar sector is characterized by 8 degrees of freedom, corresponding to the 8 integration constants that the general solution 
of \eqref{scalar eqs 1}-\eqref{scalar eqs 4} must have.

Rather surprisingly, the system \eqref{scalar eqs 1}-\eqref{scalar eqs 4} can be decoupled and its physical properties can be studied 
analytically. In order to do so, we begin by noting that we can decouple a fourth order equation for $\sA$. To obtain this equation,
we take a linear combination of \eqref{scalar eqs 1}-\eqref{scalar eqs 4} along with up to two derivatives
of \eqref{scalar eqs 1} and \eqref{scalar eqs 3}. Choosing the coefficients of this linear combination properly, we arrive at
\begin{equation}
	\sum_{i=0}^{4} b_i(y) \frac{d^i}{dy^i} \sA(y) = 0 ~,
\end{equation}
\noindent where
\begin{align}
	b_4  = 1~, \qquad 
b_3 = 4 \left(\frac{1}{y}+\frac{1}{y-1}\right) ~,
\end{align}
\begin{align}
	b_2&= \frac{2 ((y-1) \ell  (\ell +1)+y (7 y-6))}{(y-1)^2 y^2}+\frac{ \hat \omega^2 \left(2 (y-1) \epsilon ^2+2\right)}{(y-1)^2 y}~, \\
	b_1 &= \frac{2 \left(2 y+\ell ^2+\ell \right)}{(y-1)^2 y^2}+\frac{ \hat \omega^2 \left((4 y-2) \epsilon ^2+2\right)}{(y-1)^2 y^2}~, \\
\nonumber
	b_0 & = \frac{\hat \omega^4 \left((y-1) \epsilon ^2+1\right)^2}{(y-1)^4 y^2} + \frac{\ell  (\ell +1) \left(\ell ^2+\ell -2\right)}{(y-1)^2 y^4} \\
	& \frac{ \hat \omega^2  \left((y-1)^2 \epsilon ^2 \left(y+2 \left(\ell ^2+\ell +1\right)\right)+2 y \ell ^2+2 y \ell +3 y-2 \ell ^2-2 \ell -2\right)}{(y-1)^4 y^3}~.
\end{align}
The four independent solutions of this equation are of the form 
\begin{equation}\label{A gen soln}
	\sA = y^{\Delta_{\sA}} (1- y)^{- i \hat \omega} \hs _2 F_1 ( \Delta_{\sA} - i \hat \omega ( 1+\epsilon )~, 
	\Delta_{\sA} - i \hat \omega ( 1-\epsilon ) , 2 \Delta_{\sA} , y  )~,
\end{equation}
\noindent where 
\begin{equation}
	\Delta_{\sA}= \{ 2 + \ell, \ell, 1 - \ell, - 1 - \ell \}~.
\end{equation}
%
%
The associated QNM are then easily found to be
\begin{equation}\label{w A}
	\hat \omega^{(0)}_{\sA} = - \frac{i}{1\pm\epsilon} (\ell + n)~ , \qquad \hat \omega^{(2)}_{\sA} = - \frac{i}{1\pm \epsilon} (\ell +2 + n)~, \qquad n = 0, 1,2, \ldots ~. 
\end{equation}
\noindent The labels $(0)$ and $(2)$ in these quantities denote the offsets
of $0$ and $2$ with respect to $\ell$ that these modes have, respectively. Expressions \eqref{w A} are as well valid at $\epsilon =0$. 
Once a solution of ${\sA}$ is given, the remaining profiles can be found by plugging the solution back into \eqref{scalar eqs 1}-\eqref{scalar eqs 4}. While for general parameters this task can be cumbersome, it is straightforward once we set the frequencies to their QNM values 
\eqref{w A}, since then the hypergeometrics reduce to polynomials. 

The remaining four degrees of freedom can be isolated by noting that setting $\sA = 0$ we obtain a consistent set of equations 
for $\sF$, $\sS$, and $ \sAO$. To see this, we set $\sA = 0$, 
and eliminate $\sS$ algebraically from \eqref{scalar eqs 1}. By doing so, \eqref{scalar eqs 2} and \eqref{scalar eqs 3} become two coupled 
second order equations for $\sF$ and $\sAO$, while \eqref{scalar eqs 4} yields a third order equation for $\sF$ which does not provide independent information. The second order equations can be in fact decoupled introducing the fields $V_\pm$ via  
\begin{equation}\label{MF 2}
	\sF = \alpha_0(y) (V_- + V_+)~  , \qquad \sAO = \alpha_-(y) V_- + \alpha_+(y) V_+~,
\end{equation}
%
%
\begin{align}
	\alpha_0(y)&= \frac{4 \ell^2 \left((y-1) \epsilon ^2+1\right)^2+4 \ell \left((y-1) \epsilon ^2+1\right)^2+\left(\epsilon ^2-1\right) \left((y-1)^2 \epsilon ^2-1\right)}{y \left((y-1) \epsilon ^2+1\right)}~, \\
	\alpha_\pm(y) &= \frac{c_\pm \left((y-1) \epsilon ^3+\epsilon \right)+\left(\epsilon ^2-1\right)^2}{y \epsilon }~,
\end{align}
\noindent with 
\begin{equation}
	c_\pm = \frac{(\epsilon^2-1)}{\epsilon} ( 1 \pm \epsilon (2 \ell + 1) )~.
\end{equation}
The decoupled equations of motion for $V_\pm$ adopt the form
\begin{equation}
\label{eq Vpm}
	(y-1) V_\pm'' + V_\pm' + \left[  y^{-2} \frac{P_\pm(y)}{Q_{\pm}(y)^2} + \frac{(1 + (y-1)\epsilon^2)\hat \omega^2}{y(y-1)}   \right]  V_\pm   = 0 ~,
\end{equation}
\noindent where
\begin{equation}
	Q_\pm= (2 \ell +1)\epsilon^2 ( y-1 ) + (2 \ell+1) \pm \epsilon y ~,
\end{equation}
\noindent and
\begin{align}
\nonumber
	P_\pm &= \epsilon ^4 (2 \ell+1)^2 (y-1) (\ell (\ell+1) (y-1)-y) \\
\nonumber
	& \pm \epsilon ^3 (2 \ell+1) y (2 \ell (\ell+1) (1-y)-y+2) \\
\nonumber
	&+ \epsilon ^2 \left(\left(\ell^2+\ell+1\right) y^2+2 \ell (\ell+1) (2 \ell+1)^2 y-2 \ell (\ell+1) (2 \ell+1)^2\right)  \\ 
	&\mp \epsilon (2 \ell+1) y   (2 \ell (\ell+1)-y+2) + \ell (\ell+1) \left((2 \ell+1)^2-4 y\right)-y~.
\end{align}
The two second order decoupled equations \eqref{eq Vpm} capture the four degrees of freedom that we were after. We observe that the fields $V_\pm$ can be mapped into one another by the ``charge conjugation" transformation $\epsilon \to - \epsilon$. For $\epsilon = 0$, we can easily find the full solution of the system in terms of hypergeometric functions, and that the spectrum is 
given by 
%
%
%
\begin{equation}
	\hat \omega = - i (\ell + n)~, \qquad  \hat \omega = - i (\ell + 2 + n)~. 
\end{equation}
On the other hand, for any $\epsilon \neq 0$, the presence of $Q_\pm$ in \eqref{eq Vpm} modifies the structure of the singularities 
in the wave equation so it is no longer of the hypergeometric type. The ODE \eqref{eq Vpm} has four regular singular points located at $y=0,1,\infty$ and the zero of $Q_\pm$; this makes the solutions Heun functions.\footnote{Because $\ell$ is an integer, the singularity at $y=0$ is actually a resonant singularity, but it does not affect the conclusions we draw hereafter.} Because of this, we have not been able to 
solve it for general parameters, but we can show that for the frequencies
%
%
%
\begin{equation}\label{omega A=0}
	\hat \omega^{(0)}_\pm = - \frac{i}{1\pm\epsilon} (\ell + n)~ , \qquad \hat \omega^{(2)}_\pm = - \frac{i}{1\mp \epsilon} (\ell +2 + n), \qquad n = 0, 1,2, \dots , 
\end{equation}
\noindent the solutions satisfying ingoing boundary conditions and regularity at infinity can be written as polynomials of order $n$ and $n+2$, for $\hat \omega_\pm^{(0)}$ and $\hat \omega_\pm^{(2)}$ respectively. More concretely, the independent solutions take the form
\begin{align}\label{soln Vpm}
 	V_\pm^{(0)} = y^{\ell+1} (1- y)^{- i \hat \omega^{(0)}_\pm}  Q_\pm(y)^{-1}  \sum_{k=0}^{n} a^{(0)}_{k, \pm} y^k ~,\\
 	V_\pm^{(2)} = y^{\ell+1} (1- y)^{- i \hat \omega^{(2)}_\pm} Q_\pm(y)^{-1} \sum_{k=0}^{n+2} a^{(2)}_{k, \pm} y^k ~.
 \end{align} 
The coefficients $a^{(0)}_{k, \pm}$, $a^{(2)}_{k, \pm}$  can be computed order by order in $y$.\footnote{The polynomials in \eqref{soln Vpm} are formally known as a particular class of Heun polynomials. The specific form is not important, rather we want to highlight that the solution truncates.}
The profiles for the original gauge invariant fields can be easily recovered by solving the algebraic relations 
\eqref{MF 2}, and \eqref{scalar eqs 1} with $\sA=0$.
While our analysis does not {\it a priori} guarantee that all solutions of \eqref{eq Vpm} are of this form, we have checked 
this result numerically for a wide range of parameters. Moreover, continuity with the $\epsilon = 0$ case, and analogy with the
Klein-Gordon field in the subRN background, suggests that this might be the full solution of the system.  
This completes our analysis of the scalar modes with $\delta \chi =0$.

\subsubsection*{Scalar modes with $\delta \chi\neq0$}
  
Let us now consider modes with non-vanishing axion perturbation. As mentioned above, the equation of motion for the axion decouples and 
adopts the form \eqref{eq:axion}, so it can be studied on its own. In particular, this means that the QNM frequencies obtained by solving 
this equation, let's call them $\omega_{{\rm axion}}$, are QNM of the full system, despite the fact the axion does enter as a source in Maxwell's equations for $\tilde A$ and $A^0$.\footnote{Similarly to the cases discussed above, in order to obtain the profiles for the remaining fields for every $\omega_{{\rm axion}}$, we need to find the non-homogeneous solutions with these given sources.}

\begin{figure}[h]
\begin{center}
\includegraphics[scale=0.5]{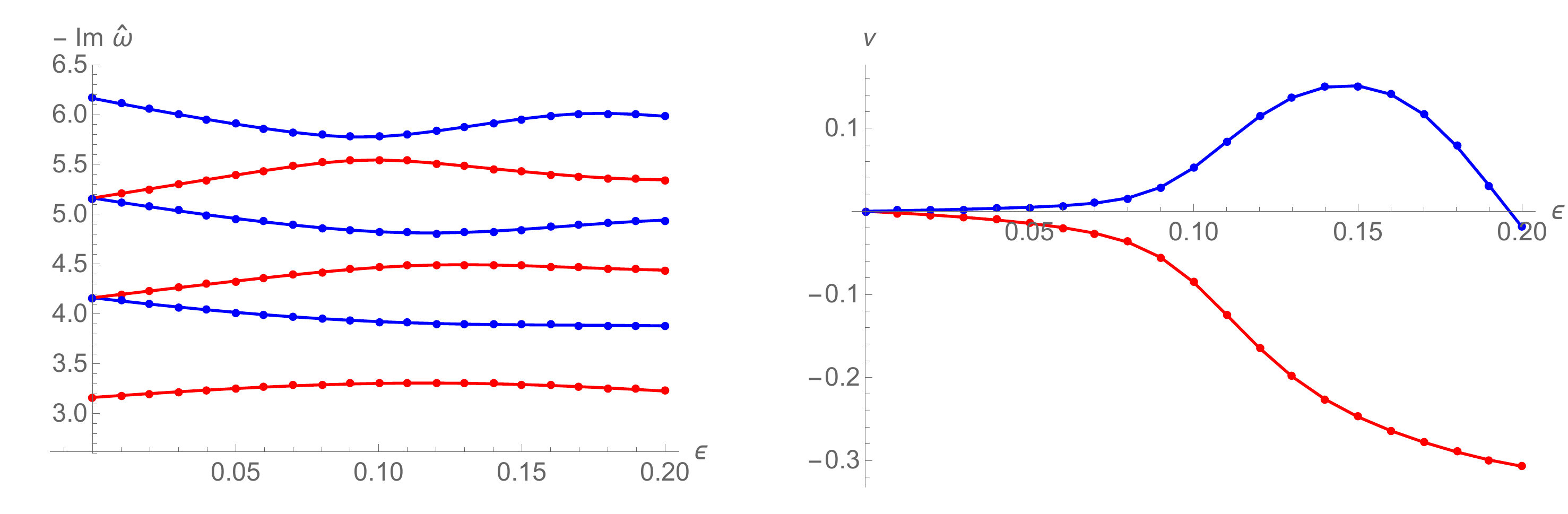}
\caption{Six lowest QNM frequencies for the axion field for $\ell = 2$. All real parts are zero and the imaginary parts are negative. On the left panel we plot $- {\rm Im} \, \hat \omega $ as a function of $\epsilon$. Modes we would expect to be evenly spaced for hypergeometric
solutions are plotted in the same color. On the right panel we display our test for even spacing defined in \eqref{nu} for $i = 0$ (red)
and $i=1$ (blue). This ceases to be zero for $\epsilon \neq 0$, so the modes are not evenly spaced.}
\label{axion QNM}
\end{center}
\end{figure}

The near boundary fall offs that follow from \eqref{eq:axion} are given by $s_2 \sim r^{\pm \delta_{{\rm axion}}}$ with 
\begin{equation}
	\delta_{{\rm axion}} =  \sqrt{\ell^2 + \ell + 4}~.
\end{equation}
The QNM are then solutions that decay as $r^{-\delta_{{\rm axion}}}$ and satisfy ingoing boundary conditions. 
As before, we perform the redefinitions \eqref{redef r w}, obtaining
\begin{equation}\label{axion eq 2}
	s_2''  + \frac{\left(y^2 \epsilon ^2-2 y \left(\epsilon ^2-1\right)+\epsilon ^2-1\right)}{(y-1) y \left((y-1) \epsilon ^2+1\right)} 
	s_2' + \left( \frac{ \delta_{{\rm axion}} (y-1) \epsilon ^2+\delta_{{\rm axion}} +2 y \epsilon }{y^2 \left((y-1)^2 \epsilon ^2+y-1\right)}  +  \frac{ \hat \omega^2 \left((y-1) \epsilon ^2+1\right)}{(y-1)^2 y} \right) s_2 = 0~.
\end{equation}
For $\epsilon = 0$, this equation can be solved in terms of hypergeometrics, and the resulting spectrum is once again 
quantized according to 
\begin{equation}
	\hat \omega = - i (\delta_{{\rm axion}} + n )~.
\end{equation} 
By studying the structure of the singularities of \eqref{axion eq 2}, we conclude that it is not a hypergeometric equation for 
$\epsilon \neq 0$; it contains four regular  singular points making the equation of the Heun type. 
While this prevents us from finding the analytic solution for general $\epsilon$, we can obtain the spectrum numerically as in the previous 
cases. All frequencies are purely imaginary, but we find that the spectrum is not evenly spaced. We plot our results in Fig. \ref{axion QNM}. In order to check for evenly spaced frequencies, we define the quantity 
\begin{equation}\label{nu}
 	\nu_n =  i ( \hat \omega_{n+4} - \hat \omega_{n+2} ) - i ( \hat \omega_{n+2} - \hat \omega_n )~.
 \end{equation} 
\noindent where $n$ denotes the overtone of the mode, $n=0$ being the lowest QNM. 
This quantity is identically zero for all $\epsilon$, $\ell$ for the frequencies \eqref{w A}. This is not the case for solutions of 
\eqref{axion eq 2} with $\epsilon \neq 0$, showing that the spectrum is not evenly spaced. 
  
%

\section*{Acknowledgements}

It is a pleasure to thank Nikos Kaplis and Ioannis Papadimitriou for helpful discussions.
TA was supported by the European Research Council under the European Union's Seventh Framework Programme 
(ERC Grant agreement 307955). 
AC is supported by Nederlandse Organisatie voor Wetenschappelijk Onderzoek (NWO) via a Vidi grant. 
The work of AC and DCM is part of the Delta ITP consortium, a program of the NWO that is funded by the Dutch Ministry of Education, Culture and Science (OCW).
TA is grateful to the Institute of Physics at University of Amsterdam 
and the Institute Lorentz at Leiden University for their hospitality during the completion of this work.
DCM would like to acknowledge the Becas Chile scholarship programme of the Government of Chile.

\appendix


\section{Equations of motion in the electric frame}\label{app:eom}

In this appendix we give the equations of motion for the electric frame STU action \eqref{eq:elecaction}. The Einstein equation is
\begin{align}
&R_{\mu\nu}-\frac12 g_{\mu\nu} R+\frac12g_{\mu\nu}\left(\frac32 \pd_\alpha\eta\pd^\alpha\eta+\frac32e^{2\eta} \pd_\alpha\chi\pd^\alpha\chi\right)- \frac32\pd_\mu\eta\pd_\nu\eta  -\frac32e^{2\eta} \pd_\mu\chi\pd_\nu\chi \cr
&+\frac12g_{\mu\nu}\left( \frac14 e^{-3\eta} (F^0)^2+ \frac{3}{4} \frac{e^{- \eta}}{(4 \chi^2 + e^{- 2 \eta})} (\tilde F - \chi^2 F^0)^2\right)\cr &-
\frac12e^{-3\eta}F^0_{\al\mu}F^{0,\al}_{\;\;\;\nu}
- \frac{3}{2} \frac{e^{- \eta}}{(4 \chi^2 + e^{- 2 \eta})} (\tilde F - \chi^2 F^0)_{\al\mu}(\tilde F - \chi^2 F^0)^{\al}_{~~\nu}=0~.
\end{align}

The equations for the dilaton and axion are 
\begin{align}\label{eq:ad}
&{1\over \sqrt{-g}}\partial_\mu(\sqrt{-g}\,\pd^\mu \eta)-e^{2\eta}\partial_\mu \chi \partial^\mu \chi+\frac14e^{-3\eta}(F^0)^2+\frac14e^{-\eta}\frac{4  \chi ^2-e^{-2\eta}}{ \left(4  \chi ^2+e^{-2\eta}\right)^2}(\tilde F-\chi^2 F^0)^2\cr
&-\frac{ 2\chi \,e^{-2 \eta } }{\left(4 \chi ^2- e^{-2 \eta }\right)^2} (\tilde F-\chi^2 F^0)\wedge (\tilde F-\chi^2 F^0)=0~,\cr
&{1\over \sqrt{-g}}\partial_\mu(\sqrt{-g}\,e^{2\eta}\pd^\mu \chi)+\frac{ \chi \,e^{-\eta}}{ \left(4  \chi ^2+e^{-2\eta}\right)^2}(\tilde F-\chi^2 F^0)_{\mu\nu} (2\tilde F+(2\chi^2+ e^{-2\eta}) F^0)^{\mu\nu} \cr
&+\frac{ 4\chi^2-e^{-2\eta}  }{\left(4 \chi ^2+ e^{-2 \eta }\right)^2} (\tilde F-\chi^2 F^0)\wedge (\tilde F-\chi^2 F^0) +{e^{-2\eta}\over 4 \chi ^2+ e^{-2 \eta }}(\tilde F-\chi^2 F^0)\wedge  F^0 =0~,
\end{align}
and the Maxwell equations are
\begin{align}
&{3\over \sqrt{-g}} \pd_\mu\le(\sqrt{-g}\,  \frac{e^{- \eta}}{(4 \chi^2 + e^{- 2 \eta})} (\tilde F-\chi^2 F^0)^{\mu\nu}\ri)\cr
&+ 3\varepsilon^{\alpha\beta\mu\nu}\tilde F_{\alpha\beta}\pd_\mu\le({\chi \over 4\chi^2+e^{-\eta}}\ri)- \varepsilon^{\alpha\beta\mu\nu} F^0_{\alpha\beta}\pd_\mu\le({\chi (2\chi^2 +e^{-2\eta})\over 4\chi^2+e^{-\eta}}\ri)=0~,\cr
 &{1\over \sqrt{-g}}\pd_\mu(\sqrt{-g}\, e^{-3\eta}F^{0,\mu\nu})+ {3\over \sqrt{-g}} \pd_\mu\le(\sqrt{-g}\,  \frac{\chi^2e^{- \eta}}{(4 \chi^2 + e^{- 2 \eta})} (\tilde F-\chi^2 F^0)^{\mu\nu}\ri)\cr
&+ 3\varepsilon^{\alpha\beta\mu\nu}\tilde F_{\alpha\beta}\pd_\mu\le({\chi (2\chi^2 +e^{-2\eta})\over 4\chi^2+e^{-\eta}}\ri)- \varepsilon^{\alpha\beta\mu\nu} F^0_{\alpha\beta}\pd_\mu\le({\chi^3 (\chi^2 +e^{-2\eta})\over 4\chi^2+e^{-\eta}}\ri)=0~,
\end{align}
In \eqref{eq:ad}, the wedge notation is short hand for a contraction with an epsilon tensor
\be
\tilde F\wedge F^0 \equiv \varepsilon^{\mu\nu\lambda \rho} \tilde F_{\mu\nu} F^0_{\lambda \rho}~,
\ee
where  $\varepsilon_{\mu\nu\lambda \rho}=\sqrt{-g}\epsilon_{\mu\nu\lambda \rho}$, and $\epsilon_{0123}=1$.
Setting $\chi=0$ simplifies these equations to 
\begin{align}
R_{\mu\nu}-\frac12 g_{\mu\nu} R-\frac12g_{\mu\nu}\left(-\frac32 \pd_\mu\eta\pd^\mu\eta-\frac14 e^{-3\eta} (F^0)^2-
\frac34 e^\eta (\tilde F)^2\right)\cr -
\frac32\pd_\mu\eta\pd_\nu\eta-
\frac12e^{-3\eta}F^0_{\al\mu}F^{0,\al}_{\;\;\;\nu}-
\frac32e^\eta \tilde F_{\al\mu}\tilde F^{\al}_{\;\;\nu}=0~,
\end{align}
and
\begin{align}
{1\over \sqrt{-g}}\partial_\mu(\sqrt{-g}\pd^\mu \eta)+\frac14(F^0)^2e^{-3\eta}
-\frac14 \tilde F^2 e^\eta=0~,\cr
\pd_\mu(\sqrt{-g}\,e^\eta \tilde F^{\mu\nu})=0~,\qquad \pd_\mu(\sqrt{-g}\, e^{-3\eta}F^{0,\mu\nu})=0~,
\end{align}
with the addition of the constraint
\be
\tilde F \wedge \tilde F - e^{- 2 \eta} F^0 \wedge \tilde F = 0~.
\ee

\section{Details of the scalar mode calculation}\label{app:coef}

The coefficients $\al_i(r)$ that participate in the definitions \eqref{MF1}-\eqref{MF4} are given by
\bea
\al_1&=&\frac{4im^2(\Pi_c^2-\Pi_s^2)\omega\De^{3/2}X'}{k_S^2\Pi_c\Pi_sX}
-\frac{4im(\Pi_c^2-\Pi_s^2)\omega\De^{5/2}}{k_S^2\Pi_c\Pi_sX\De'}-\\
&-&\frac{4im^2\omega(4m^2\Pi_c^2\Pi_s^2-(\Pi_c^2-\Pi_s^2)^2X)\sqrt\De\,\De'}{k_S^2\Pi_c\Pi_s(\Pi_c^2-\Pi_s^2)X},
\nonumber\\
\al_2&=&\frac{m^2(\Pi_c^2-\Pi_s^2)\sqrt\De(X+2\omega^2\De)X'}{k_S^3\Pi_c\Pi_sX}-
\frac{2m^2(\Pi_c^2-\Pi_s^2)\omega^2\De^{5/2}}{k_S^3\Pi_c\Pi_sX\De'}-m^2\De'\times\\
&\times&
\frac{7(\Pi_c^2-\Pi_s^2)^2X^2-16m^2\Pi_c^2\Pi_s^2\omega^2\De+
2X(8k_S^2m^2\Pi_c^2\Pi_s^2+(\Pi_c^2-\Pi_s^2)^2(2\omega^2\De+X'^2))}{2k_S^3\Pi_c\Pi_s(\Pi_c^2-\Pi_s^2)X\sqrt\De}\nonumber+\\
&+&\frac{m^2(8m^2\Pi_c^2\Pi_s^2-9(\Pi_c^2-\Pi_s^2)^2X)X'\De'^2}{2k_S^3\Pi_c\Pi_s(\Pi_c^2-\Pi_s^2)\De^{3/2}}+
\frac{7m^2X(4m^2\Pi_c^2\Pi_s^2-(\Pi_c^2-\Pi_s^2)X)\De'^3}{2k_S^3\Pi_c\Pi_s(\Pi_c^2-\Pi_s^2)\De^{5/2}},\\
\al_3&=&\frac{2m^2(\Pi_c^2-\Pi_s^2)\sqrt\De(3X+X'^2)}{k_S^3\Pi_c\Pi_s}-
\frac{2m^2(\Pi_c^2-\Pi_s^2)\De^{3/2}X'}{k_S^3\Pi_c\Pi_s\De'}+\\
&+&\frac{8m^2(m^2\Pi_c^2\Pi_s^2-(\Pi_c^2-\Pi_s^2)^2X)X'\De'}{k_S^3\Pi_c\Pi_s(\Pi_c^2-\Pi_s^2)\sqrt\De}-\frac{6m^2X(4m^2\Pi_c^2\Pi_s^2-(\Pi_c^2-\Pi_s^2)^2X)\De'^2}{k_S^3\Pi_c\Pi_s(\Pi_c^2-\Pi_s^2)\De^{3/2}}
\nonumber,\\
\al_4&=&\frac{2m^2(\Pi_c^2-\Pi_s^2)X\sqrt\De X'}{k_S^3\Pi_c\Pi_s}-
\frac{2m^2(\Pi_c^2-\Pi_s^2)X\De^{3/2}}{k_S^3\Pi_c\Pi_s\De'}+\\
&+&\frac{2m^2X(4m^2\Pi_c^2\Pi_s^2-(\Pi_c^2-\Pi_s^2)^2X)\De'}{k_S^3\Pi_c\Pi_s(\Pi_c^2-\Pi_s^2)\sqrt\De}.\nonumber
\eea
The coefficients that appear in the equations of motion for the linearized perturbations in the scalar sector 
\eqref{scalar eqs 1}-\eqref{scalar eqs 4} are
\bea
c_{1\hat{\mc A}'}&=&1/(-1 + y),\\
c_{1\mc A}&=&\frac{4 \hat{\omega} ^2 \left((y-1) \epsilon ^2+1\right)}{(y-1)^2 y}+
\frac{k_S^2}{(y-1) y^2}\\
&-&\frac{3 \left(\epsilon ^2-1\right)^2}{y^2 \left(8 k_S^2 \left((y-1) \epsilon ^2+1\right)^2+\left(\epsilon ^2-1\right) \left((y-1) (2 y-5) \epsilon ^2+3 y-5\right)\right)}\nonumber,\\
c_{1\hat\Phi}&=&-\frac{\left(\epsilon ^2-1\right) k_S^3 \left((y-1) \epsilon ^2+1\right)}{2 (y-1) y \left(8 k_S^2 \left((y-1) \epsilon ^2+1\right)^2+\left(\epsilon ^2-1\right) \left((y-1) (2 y-5) \epsilon ^2+3 y-5\right)\right)},\\
c_{1\hat\Phi'}&=&\frac{\left(\epsilon ^2-1\right)^2 k_S}{4 \left(8 k_S^2 \left((y-1) \epsilon ^2+1\right)^2+\left(\epsilon ^2-1\right) \left((y-1) (2 y-5) \epsilon ^2+3 y-5\right)\right)},\\
c_{1\hat S}&=&\left[
4 k_S^2 \left((y-1) \epsilon ^2+1\right)^2-\left(1-\epsilon ^2\right) \left((y-1)^2 \epsilon ^2-1\right)
\right]\times\left[
24 (1-y)^{3/2} y \left(\epsilon ^2-1\right)
\right.\nonumber\\
&\times&\left.8 k_S^2 \left((y-1) \epsilon ^2+1\right)^2+\left(\epsilon ^2-1\right) \left(2 y^2 \epsilon ^2-7 y \epsilon ^2+3 y+5 \epsilon ^2-5\right)\nonumber\right]^{-1},\\
c_{1\hat{\mc A_0}}&=&-\epsilon  \left(\epsilon ^2-1\right)^2\\
&\times&\left[256 \left((y-1) \epsilon ^2+1\right)^2 \left(8 k_S^2 \left((y-1) \epsilon ^2+1\right)^2+\left(\epsilon ^2-1\right) \left((y-1) (2 y-5) \epsilon ^2+3 y-5\right)\right)\right]^{-1}
\nonumber,\\
c_{2\hat{\mc A'_0}}&=&\frac{(y-2) (y-1) \epsilon ^2+3 y-2}{(y-1) y \left((y-1) \epsilon ^2+1\right)},\\
c_{2\hat{\mc A_0}}&=&\frac{k_S^2}{(y-1) y^2}+\frac{4 \hat{\omega} ^2 \left((y-1) \epsilon ^2+1\right)}{(y-1)^2 y}\\
&-&\frac{\epsilon ^2 \left(\epsilon ^2-1\right)^2}{\left((y-1) \epsilon ^2+1\right)^2 \left(8 k_S^2 \left((y-1) \epsilon ^2+1\right)^2+\left(\epsilon ^2-1\right) \left((y-1) (2 y-5) \epsilon ^2+3 y-5\right)\right)}
\nonumber,\\
c_{2\hat\Phi}&=&-\frac{128 \epsilon  \left(\epsilon ^2-1\right) k_S^3 \left((y-1) \epsilon ^2+1\right)}{(y-1) y \left(8 k_S^2 \left((y-1) \epsilon ^2+1\right)^2+\left(\epsilon ^2-1\right) \left((y-1) (2 y-5) \epsilon ^2+3 y-5\right)\right)},\\
c_{2\hat\Phi'}&=&\frac{64 \epsilon  \left(\epsilon ^2-1\right)^2 k_S}{8 k_S^2 \left((y-1) \epsilon ^2+1\right)^2+\left(\epsilon ^2-1\right) \left(2 y^2 \epsilon ^2-7 y \epsilon ^2+3 y+5 \epsilon ^2-5\right)},\\
c_{2\hat S}&=&\left[
-32 \epsilon  \left(4 k_S^2 \left((y-1) \epsilon ^2+1\right)^2+\left(\epsilon ^2-1\right) \left((y-3) (y-1) \epsilon ^2+2 y-3\right)\right)
\right]\\
&\times&\left[
(1-y)^{3/2} y \left(\epsilon ^2-1\right)
\right.
\nonumber\\
&\times&\left.
8 k_S^2 \left((y-1) \epsilon ^2+1\right)^2+\left(\epsilon ^2-1\right) \left((y-1) (2 y-5) \epsilon ^2+3 y-5\right)\right]^{-1}
\nonumber,\\
c_{2\hat{\mc A_0}}&=&-\frac{768 \epsilon  \left(\epsilon ^2-1\right)^2}{y^2 \left(8 k_S^2 \left((y-1) \epsilon ^2+1\right)^2+\left(\epsilon ^2-1\right) \left(2 y^2 \epsilon ^2-7 y \epsilon ^2+3 y+5 \epsilon ^2-5\right)\right)},
\eea
\bea
c_{3\hat\Phi'}&=&\frac{(y-2) (y-1) \epsilon ^2+3 y-2}{(y-1) y \left((y-1) \epsilon ^2+1\right)}\\
&-&\frac{\left(\epsilon ^2-1\right) \left(3 (y-1) (2 y-1) \epsilon ^4+2 (y (y+3)-3) \epsilon ^2+3 (y+1)\right)}{y \left((y-1) \epsilon ^2+1\right) \left(8 k_S^2 \left((y-1) \epsilon ^2+1\right)^2+\left(\epsilon ^2-1\right) \left((y-1) (2 y-5) \epsilon ^2+3 y-5\right)\right)}
\nonumber,\\
c_{3\hat\Phi}&=&\frac{\epsilon ^2 \left(7 \epsilon ^2+1\right)}{4 (y-1) \left(y \epsilon ^2-\epsilon ^2+1\right)^2}-\frac{(\epsilon -1) (\epsilon +1) \left(25 \epsilon ^2+3\right)}{8 (y-1) y \left(y \epsilon ^2-\epsilon ^2+1\right)^2}\\
&-&\frac{k_S^2}{(y-1) y^2}+\frac{4 \hat{\omega} ^2 \left((y-1) \epsilon ^2+1\right)}{(y-1)^2 y}+\frac{\left(\epsilon ^2-1\right) \left((y (2 y-7)-6) \epsilon ^2+3 (y+2)\right)}{8 (y-1) \left((y-1) y \epsilon ^2+y\right)^2}
\nonumber\\
&-&\frac{\left(\epsilon ^2-1\right) \left((y-1) (2 y-5) \epsilon ^2+3 y-5\right) \left(3 (y-1) (2 y-1) \epsilon ^4+2 (y (y+3)-3) \epsilon ^2+3 (y+1)\right)}{4 (y-1) \left((y-1) y \epsilon ^2+y\right)^2 \left(8 k_S^2 \left((y-1) \epsilon ^2+1\right)^2+\left(\epsilon ^2-1\right) \left((y-1) (2 y-5) \epsilon ^2+3 y-5\right)\right)}\nonumber
\eea
\bea
c_{3\hat S}&=&\left[
(y-1) (2 (y-3) (y-1) y-3) \epsilon ^4+(y (y (3 y-5)+3)-3) \epsilon ^2+y+1
\right.\\
&-&\left.4 k_S^2 \left((y-1) \epsilon ^2+1\right)^2 \left((y-1) (2 y-1) \epsilon ^2-y-1\right)-(y-1)^3 (2 y-1) \epsilon ^6\right]
\nonumber\\
&\times&\left[
2 (1-y)^{3/2} y^2 \left(\epsilon ^2-1\right) k_S \left((y-1) \epsilon ^2+1\right)
\right.
\nonumber\\
&\times&\left.
8 k_S^2 \left((y-1) \epsilon ^2+1\right)^2+\left(\epsilon ^2-1\right) \left((y-1) (2 y-5) \epsilon ^2+3 y-5\right)\right]^{-1}\nonumber,\\
c_{3\hat{\mc A}}&=&-\frac{48 k_S \left((y-1) \epsilon ^2+1\right)}{(y-1) y^3 \left(\epsilon ^2-1\right)}\\
&+&\frac{12 \left(3 \left(y \left(y^2+y-4\right)+2\right) \epsilon ^2+(y-1)^3 (2 y+3) \epsilon ^4+5 y-3\right)}{(y-1) y^3 k_S \left((y-1)^2 (2 y-5) \epsilon ^4+5 (y-2) (y-1) \epsilon ^2+3 y-5\right)}\nonumber\\
&+&\frac{96 k_S \left((y-1) \epsilon ^2+1\right) \left(3 (y-1) (2 y-1) \epsilon ^4+2 (y (y+3)-3) \epsilon ^2+3 (y+1)\right)}{y^3 \left((y-1) (2 y-5) \epsilon ^2+3 y-5\right) \left(8 k_S^2 \left((y-1) \epsilon ^2+1\right)^2+\left(\epsilon ^2-1\right) \left((y-1) (2 y-5) \epsilon ^2+3 y-5\right)\right)}\nonumber\\
c_{3\hat{\mc A_0}}&=&-\frac{\epsilon  \left(\epsilon ^2-1\right)^3 k_S}{16 (y-1) y \left((y-1) \epsilon ^2+1\right)^5}\\
&-&\frac{y^3 \epsilon ^9 k_S}{16 (y-1) \left(\epsilon ^2-1\right) \left((y-1) \epsilon ^2+1\right)^5}
+\frac{\epsilon ^3 k_S \left((y (2 y-3)+2) \epsilon ^4+(3 y-4) \epsilon ^2+2\right)}{8 (y-1) \left((y-1) \epsilon ^2+1\right)^5}
\nonumber\\
&+&\frac{\epsilon  \left(-6 y^2+(y-1)^2 (y (2 y+5)-13) \epsilon ^4-(y-1) ((y-26) y+26) \epsilon ^2+21 y-13\right)}{64 (y-1) y k_S \left((y-1) \epsilon ^2+1\right)^3 \left((y-1) (2 y-5) \epsilon ^2+3 y-5\right)}
\nonumber\\
&-&\left[\epsilon  k_S \left(3 (y-1) (2 y-1) \epsilon ^4+2 (y (y+3)-3) \epsilon ^2+3 (y+1)\right)\right]\nonumber\\
&\times&\left[
8 y \left((y-1) \epsilon ^2+1\right) \left((y-1) (2 y-5) \epsilon ^2+3 y-5\right)
\right.\nonumber\\
&\times&\left.
8 k_S^2 \left((y-1) \epsilon ^2+1\right)^2+\left(\epsilon ^2-1\right) \left((y-1) (2 y-5) \epsilon ^2+3 y-5\right)
\right]^{-1}\nonumber,
\eea
\bea
c_{4\hat S'}&=&\frac2y\\
&-&\frac{3 (y-1) \left(\epsilon ^2-1\right)^3}{y \left((y-1) \epsilon ^2+1\right) \left(8 k_S^2 \left((y-1) \epsilon ^2+1\right)^2+\left(\epsilon ^2-1\right) \left((y-1) (2 y-5) \epsilon ^2+3 y-5\right)\right)}
\nonumber,\\
c_{4\hat S}&=&-\frac{\left(\epsilon ^2-1\right) \left((y-1) (2 y-5) \epsilon ^2+3 y-5\right)}{4 (y-1) \left((y-1) y \epsilon ^2+y\right)^2}\\
&+&\frac{(y-1)^2 (y (y+4)-10) \epsilon ^4+2 y (y+4) (3 y-5) \epsilon ^2+y (16-5 y)+20 \epsilon ^2-10}{4 (y-1)^2 y^2 \left((y-1) \epsilon ^2+1\right)^2}
\nonumber\\
&+&\frac{k_S^2}{(y-1) y^2}+\frac{4 \hat{\omega} ^2 \left((y-1) \epsilon ^2+1\right)}{(y-1)^2 y}
\nonumber\\
&-&\frac{3 (y-1) \left(\epsilon ^2-1\right)^4 \left(3 (y-1)^2 \epsilon ^4+(y (y+6)-6) \epsilon ^2+3\right)}{y^2 \left((y-1) \epsilon ^2+1\right)^2 \left(8 k_S^2 \left((y-1) \epsilon ^2+1\right)^2+\left(\epsilon ^2-1\right) \left((y-1) (2 y-5) \epsilon ^2+3 y-5\right)\right){}^2}
\nonumber\\
&-&\frac{3 \left(\epsilon ^2-1\right)^3 \left((y-1) (2 y-1) \epsilon ^2-y-1\right)}{4 \left((y-1) y \epsilon ^2+y\right)^2 \left(8 k_S^2 \left((y-1) \epsilon ^2+1\right)^2+\left(\epsilon ^2-1\right) \left((y-1) (2 y-5) \epsilon ^2+3 y-5\right)\right)},
\nonumber\\
c_{4\hat\Phi}&=&6 \left(\epsilon ^2-1\right)^3 k_S\\
&\times&\left[
8 y \hat{\omega} ^2 \left(\epsilon ^2-1\right) \left(8 k_S^2 \left((y-1) \epsilon ^2+1\right)^2+\left(\epsilon ^2-1\right) \left((y-1) (2 y-5) \epsilon ^2+3 y-5\right)\right)
\right.\nonumber\\
&-&\frac{k_S^2}{(y-1) \epsilon ^2+1}\nonumber\\
&\times&\left(\epsilon ^2-1\right) \left((y-1)^2 \left(8 y^2-42 y+37\right) \epsilon ^4+3 y^2+2 (y-1) (y (5 y-37)+37) \epsilon ^2-32 y+37\right)
\nonumber\\
&+&\left.\left.8 k_S^2 \left((y-1) \epsilon ^2+1\right)^2 \left((y-1) (4 y-5) \epsilon ^2+y-5\right)\right)\right]
\nonumber\\
&\times&\left[
\sqrt{1-y} y^2 \left(8 k_S^2 \left((y-1) \epsilon ^2+1\right)^2+\left(\epsilon ^2-1\right) \left((y-1) (2 y-5) \epsilon ^2+3 y-5\right)\right){}^2
\right]^{-1},\nonumber\\
c_{4\hat\Phi'}&=&-6 \sqrt{1-y} \left(\epsilon ^2-1\right)^4 k_S\\
&\times&\left[
\left(\epsilon ^2-1\right) \left((y-1)^3 (2 y-11) \epsilon ^4+(y-1) ((y-22) y+22) \epsilon ^2-9 y+11\right)
\right.
\nonumber\\
&+&\left.8 k_S^2 \left((y-1) \epsilon ^2+1\right)^2 \left((y-1)^2 \epsilon ^2-1\right)\right]\nonumber\\
&\times&\left[
y \left((y-1) \epsilon ^2+1\right)^2 \left(8 k_S^2 \left((y-1) \epsilon ^2+1\right)^2+\left(\epsilon ^2-1\right) \left((y-1) (2 y-5) \epsilon ^2+3 y-5\right)\right){}^2
\right]^{-1}\nonumber,\\
c_{4\hat{\mc A}}&=&24 \left(\epsilon ^2-1\right)\\
&\times&\left[
-3 (y-1) \left(\epsilon ^2-1\right)^4 \left((y-1)^3 (2 y-11) \epsilon ^4+(y-1) ((y-22) y+22) \epsilon ^2-9 y+11\right)
\right.\nonumber\\
&+&2 \left(\epsilon ^2-1\right)^2 \left((y-1) \epsilon ^2+1\right)^2\nonumber\\
&\times&k_S^2 \left(27 y^2+(y-1)^2 (4 (y-11) y+67) \epsilon ^4+2 (y-1) (y (18 y-67)+67) \epsilon ^2-90 y+67\right)
\nonumber\\
&+&\left.\left.128 k_S^6 \left((y-1) \epsilon ^2+1\right)^6++64 \left(\epsilon ^2-1\right) k_S^4 \left((y-1) \epsilon ^2+1\right)^4 \left((y-4) (y-1) \epsilon ^2+3 y-4\right)\right)\right]\nonumber\\
&\times&\left[
\sqrt{1-y} y^3 \left((y-1) \epsilon ^2+1\right)^2 \left(8 k_S^2 \left((y-1) \epsilon ^2+1\right)^2+\left(\epsilon ^2-1\right) \left((y-1) (2 y-5) \epsilon ^2+3 y-5\right)\right){}^2
\right]^{-1}\nonumber,
\eea
\bea
c_{4\hat{\mc A'}}&=&-\frac{72 \sqrt{1-y} \left(\epsilon ^2-1\right)^2 \left(8 k_S^2 \left((y-1) \epsilon ^2+1\right)^2+\left(\epsilon ^2-1\right) \left((y-1) (2 y-3) \epsilon ^2+y-3\right)\right)}{y^2 \left((y-1) \epsilon ^2+1\right) \left(8 k_S^2 \left((y-1) \epsilon ^2+1\right)^2+\left(\epsilon ^2-1\right) \left((y-1) (2 y-5) \epsilon ^2+3 y-5\right)\right)},\\
c_{4\hat{\mc A_0'}}&=&-\frac{3 \sqrt{1-y} \epsilon  \left(\epsilon ^2-1\right)^2 \left(8 k_S^2 \left((y-1) \epsilon ^2+1\right)^2+\left(\epsilon ^2-1\right) \left(2 y^2 \epsilon ^2-5 y \epsilon ^2+y+3 \epsilon ^2-3\right)\right)}{32 \left((y-1) \epsilon ^2+1\right)^3 \left(8 k_S^2 \left((y-1) \epsilon ^2+1\right)^2+\left(\epsilon ^2-1\right) \left(2 y^2 \epsilon ^2-7 y \epsilon ^2+3 y+5 \epsilon ^2-5\right)\right)},\\
c_{4\hat{\mc A_0}}&=&-3 \epsilon  \left(\epsilon ^2-1\right)\\
&\times&\left[
(y-1) \left(\epsilon ^2-1\right)^4 \left((y-1)^3 (2 y-11) \epsilon ^4+(y-1) ((y-22) y+22) \epsilon ^2-9 y+11\right)
\right.\nonumber\\
&+&2 \left(\epsilon ^2-1\right)^2 \left((y-1) \epsilon ^2+1\right)^2\nonumber\\
&\times&(3 y^2+(y-1)^2 (4 (y-3) y+11) \epsilon ^4+2 (y-1) (y (2 y-11)+11) \epsilon ^2-10 y+11)k_S^2\nonumber\\
&+&64 (y-2) \left(\epsilon ^2-1\right) k_S^4 \left((y-1) \epsilon ^2+1\right)^5\nonumber\\
&+&\left.128 k_S^6 \left((y-1) \epsilon ^2+1\right)^6\right]\nonumber\\
&\times&\left[
\sqrt{4-4 y} y \left(2 (y-1) \epsilon ^2+2\right)^4 \left(8 k_S^2 \left((y-1) \epsilon ^2+1\right)^2+\left(\epsilon ^2-1\right) \left((y-1) (2 y-5) \epsilon ^2+3 y-5\right)\right){}^2
\right]^{-1}\nonumber
\eea

\bibliographystyle{JHEP}
\bibliography{refs}

\providecommand{\href}[2]{#2}\begingroup\raggedright\begin{thebibliography}{10}

\bibitem{FerraraKalloshStrominger1995}
S.~Ferrara, R.~Kallosh and A.~Strominger, \emph{{N=2 extremal black holes}},
  \href{http://dx.doi.org/10.1103/PhysRevD.52.R5412}{\emph{Phys. Rev.} {\bf
  D52} (1995) R5412--R5416}, [\href{https://arxiv.org/abs/hep-th/9508072}{{\tt
  hep-th/9508072}}].

\bibitem{FerraraKallosh1996}
S.~Ferrara and R.~Kallosh, \emph{{Supersymmetry and attractors}},
  \href{http://dx.doi.org/10.1103/PhysRevD.54.1514}{\emph{Phys. Rev.} {\bf D54}
  (1996) 1514--1524}, [\href{https://arxiv.org/abs/hep-th/9602136}{{\tt
  hep-th/9602136}}].

\bibitem{FerraraKallosh1996a}
S.~Ferrara and R.~Kallosh, \emph{{Universality of supersymmetric attractors}},
  \href{http://dx.doi.org/10.1103/PhysRevD.54.1525}{\emph{Phys. Rev.} {\bf D54}
  (1996) 1525--1534}, [\href{https://arxiv.org/abs/hep-th/9603090}{{\tt
  hep-th/9603090}}].

\bibitem{GuicaHartmanSongEtAl2009}
M.~Guica, T.~Hartman, W.~Song and A.~Strominger, \emph{{The Kerr/CFT
  Correspondence}},
  \href{http://dx.doi.org/10.1103/PhysRevD.80.124008}{\emph{Phys. Rev.} {\bf
  D80} (2009) 124008}, [\href{https://arxiv.org/abs/0809.4266}{{\tt
  0809.4266}}].

\bibitem{Castro:2010fd}
A.~Castro, A.~Maloney and A.~Strominger, \emph{{Hidden Conformal Symmetry of
  the Kerr Black Hole}},
  \href{http://dx.doi.org/10.1103/PhysRevD.82.024008}{\emph{Phys. Rev.} {\bf
  D82} (2010) 024008}, [\href{https://arxiv.org/abs/1004.0996}{{\tt
  1004.0996}}].

\bibitem{MaldacenaStrominger1998a}
J.~M. Maldacena and A.~Strominger, \emph{{AdS(3) black holes and a stringy
  exclusion principle}},
  \href{http://dx.doi.org/10.1088/1126-6708/1998/12/005}{\emph{JHEP} {\bf 12}
  (1998) 005}, [\href{https://arxiv.org/abs/hep-th/9804085}{{\tt
  hep-th/9804085}}].

\bibitem{BalasubramanianKrausLawrence1999}
V.~Balasubramanian, P.~Kraus and A.~E. Lawrence, \emph{{Bulk versus boundary
  dynamics in anti-de Sitter space-time}},
  \href{http://dx.doi.org/10.1103/PhysRevD.59.046003}{\emph{Phys. Rev.} {\bf
  D59} (1999) 046003}, [\href{https://arxiv.org/abs/hep-th/9805171}{{\tt
  hep-th/9805171}}].

\bibitem{ChenXueZhang2013}
B.~Chen, Z.~Xue and J.-J. Zhang, \emph{{Note on Thermodynamic Method of Black
  Hole/CFT Correspondence}},
  \href{http://dx.doi.org/10.1007/JHEP03(2013)102}{\emph{JHEP} {\bf 03} (2013)
  102}, [\href{https://arxiv.org/abs/1301.0429}{{\tt 1301.0429}}].

\bibitem{ChenLiuZhang2012}
B.~Chen, S.-x. Liu and J.-j. Zhang, \emph{{Thermodynamics of Black Hole
  Horizons and Kerr/CFT Correspondence}},
  \href{http://dx.doi.org/10.1007/JHEP11(2012)017}{\emph{JHEP} {\bf 11} (2012)
  017}, [\href{https://arxiv.org/abs/1206.2015}{{\tt 1206.2015}}].

\bibitem{CastroLapanMaloneyEtAl2013a}
A.~Castro, J.~M. Lapan, A.~Maloney and M.~J. Rodriguez, \emph{{Black Hole
  Monodromy and Conformal Field Theory}},
  \href{http://dx.doi.org/10.1103/PhysRevD.88.044003}{\emph{Phys. Rev.} {\bf
  D88} (2013) 044003}, [\href{https://arxiv.org/abs/1303.0759}{{\tt
  1303.0759}}].

\bibitem{CastroLapanMaloneyEtAl2013}
A.~Castro, J.~M. Lapan, A.~Maloney and M.~J. Rodriguez, \emph{{Black Hole
  Scattering from Monodromy}},
  \href{http://dx.doi.org/10.1088/0264-9381/30/16/165005}{\emph{Class. Quant.
  Grav.} {\bf 30} (2013) 165005}, [\href{https://arxiv.org/abs/1304.3781}{{\tt
  1304.3781}}].

\bibitem{Cvetic:2011hp}
M.~Cvetic and F.~Larsen, \emph{{Conformal Symmetry for General Black Holes}},
  \href{http://dx.doi.org/10.1007/JHEP02(2012)122}{\emph{JHEP} {\bf 02} (2012)
  122}, [\href{https://arxiv.org/abs/1106.3341}{{\tt 1106.3341}}].

\bibitem{Cvetic:2011dn}
M.~Cvetic and F.~Larsen, \emph{{Conformal Symmetry for Black Holes in Four
  Dimensions}}, \href{http://dx.doi.org/10.1007/JHEP09(2012)076}{\emph{JHEP}
  {\bf 09} (2012) 076}, [\href{https://arxiv.org/abs/1112.4846}{{\tt
  1112.4846}}].

\bibitem{Virmani:2012kw}
A.~Virmani, \emph{{Subtracted Geometry From Harrison Transformations}},
  \href{http://dx.doi.org/10.1007/JHEP07(2012)086}{\emph{JHEP} {\bf 07} (2012)
  086}, [\href{https://arxiv.org/abs/1203.5088}{{\tt 1203.5088}}].

\bibitem{Sahay:2013xda}
A.~Sahay and A.~Virmani, \emph{{Subtracted Geometry from Harrison
  Transformations: II}},
  \href{http://dx.doi.org/10.1007/JHEP07(2013)089}{\emph{JHEP} {\bf 07} (2013)
  089}, [\href{https://arxiv.org/abs/1305.2800}{{\tt 1305.2800}}].

\bibitem{Cvetic:2013vqi}
M.~Cvetic, M.~Guica and Z.~H. Saleem, \emph{{General black holes, untwisted}},
  \href{http://dx.doi.org/10.1007/JHEP09(2013)017}{\emph{JHEP} {\bf 09} (2013)
  017}, [\href{https://arxiv.org/abs/1302.7032}{{\tt 1302.7032}}].

\bibitem{Cvetic:2012tr}
M.~Cvetic and G.~W. Gibbons, \emph{{Conformal Symmetry of a Black Hole as a
  Scaling Limit: A Black Hole in an Asymptotically Conical Box}},
  \href{http://dx.doi.org/10.1007/JHEP07(2012)014}{\emph{JHEP} {\bf 07} (2012)
  014}, [\href{https://arxiv.org/abs/1201.0601}{{\tt 1201.0601}}].

\bibitem{Baggio:2012db}
M.~Baggio, J.~de~Boer, J.~I. Jottar and D.~R. Mayerson, \emph{{Conformal
  Symmetry for Black Holes in Four Dimensions and Irrelevant Deformations}},
  \href{http://dx.doi.org/10.1007/JHEP04(2013)084}{\emph{JHEP} {\bf 04} (2013)
  084}, [\href{https://arxiv.org/abs/1210.7695}{{\tt 1210.7695}}].

\bibitem{Cvetic:2014nta}
M.~Cvetic, G.~W. Gibbons and Z.~H. Saleem, \emph{{Thermodynamics of
  Asymptotically Conical Geometries}},
  \href{http://dx.doi.org/10.1103/PhysRevLett.114.231301}{\emph{Phys. Rev.
  Lett.} {\bf 114} (2015) 231301}, [\href{https://arxiv.org/abs/1412.5996}{{\tt
  1412.5996}}].

\bibitem{An:2016fzu}
O.~S. An, M.~Cvetic and I.~Papadimitriou, \emph{{Black hole thermodynamics from
  a variational principle: Asymptotically conical backgrounds}},
  \href{http://dx.doi.org/10.1007/JHEP03(2016)086}{\emph{JHEP} {\bf 03} (2016)
  086}, [\href{https://arxiv.org/abs/1602.01508}{{\tt 1602.01508}}].

\bibitem{Cvetic:2014ina}
M.~Cvetic, G.~W. Gibbons and Z.~H. Saleem, \emph{{Quasinormal modes for
  subtracted rotating and magnetized geometries}},
  \href{http://dx.doi.org/10.1103/PhysRevD.90.124046}{\emph{Phys. Rev.} {\bf
  D90} (2014) 124046}, [\href{https://arxiv.org/abs/1401.0544}{{\tt
  1401.0544}}].

\bibitem{Cvetic:2014eka}
M.~Cvetic, G.~W. Gibbons, Z.~H. Saleem and A.~Satz, \emph{{Vacuum Polarization
  of STU Black Holes and their Subtracted Geometry Limit}},
  \href{http://dx.doi.org/10.1007/JHEP01(2015)130}{\emph{JHEP} {\bf 01} (2015)
  130}, [\href{https://arxiv.org/abs/1411.4658}{{\tt 1411.4658}}].

\bibitem{Chakraborty:2012nu}
A.~Chakraborty and C.~Krishnan, \emph{{Subttractors}},
  \href{http://dx.doi.org/10.1007/JHEP08(2013)057}{\emph{JHEP} {\bf 08} (2013)
  057}, [\href{https://arxiv.org/abs/1212.1875}{{\tt 1212.1875}}].

\bibitem{Chakraborty:2012fx}
A.~Chakraborty and C.~Krishnan, \emph{{Attraction, with Boundaries}},
  \href{http://dx.doi.org/10.1088/0264-9381/31/4/045009}{\emph{Class. Quant.
  Grav.} {\bf 31} (2014) 045009}, [\href{https://arxiv.org/abs/1212.6919}{{\tt
  1212.6919}}].

\bibitem{Andrade:2014kba}
T.~Andrade, C.~Keeler, A.~Peach and S.~F. Ross, \emph{{Schrödinger holography
  with z = 2}},
  \href{http://dx.doi.org/10.1088/0264-9381/32/8/085006}{\emph{Class. Quant.
  Grav.} {\bf 32} (2015) 085006}, [\href{https://arxiv.org/abs/1412.0031}{{\tt
  1412.0031}}].

\bibitem{Chemissany:2014xsa}
W.~Chemissany and I.~Papadimitriou, \emph{{Lifshitz holography: The whole
  shebang}}, \href{http://dx.doi.org/10.1007/JHEP01(2015)052}{\emph{JHEP} {\bf
  01} (2015) 052}, [\href{https://arxiv.org/abs/1408.0795}{{\tt 1408.0795}}].

\bibitem{Kodama:2003jz}
H.~Kodama and A.~Ishibashi, \emph{{A Master equation for gravitational
  perturbations of maximally symmetric black holes in higher dimensions}},
  \href{http://dx.doi.org/10.1143/PTP.110.701}{\emph{Prog. Theor. Phys.} {\bf
  110} (2003) 701--722}, [\href{https://arxiv.org/abs/hep-th/0305147}{{\tt
  hep-th/0305147}}].

\bibitem{Kodama:2003kk}
H.~Kodama and A.~Ishibashi, \emph{{Master equations for perturbations of
  generalized static black holes with charge in higher dimensions}},
  \href{http://dx.doi.org/10.1143/PTP.111.29}{\emph{Prog. Theor. Phys.} {\bf
  111} (2004) 29--73}, [\href{https://arxiv.org/abs/hep-th/0308128}{{\tt
  hep-th/0308128}}].

\bibitem{BirminghamSachsSolodukhin2002}
D.~Birmingham, I.~Sachs and S.~N. Solodukhin, \emph{{Conformal field theory
  interpretation of black hole quasinormal modes}},
  \href{http://dx.doi.org/10.1103/PhysRevLett.88.151301}{\emph{Phys. Rev.
  Lett.} {\bf 88} (2002) 151301},
  [\href{https://arxiv.org/abs/hep-th/0112055}{{\tt hep-th/0112055}}].

\bibitem{Cvetic:2016eiv}
M.~Cvetic and I.~Papadimitriou, \emph{{AdS$_2$ Holographic Dictionary}},
  \href{https://arxiv.org/abs/1608.07018}{{\tt 1608.07018}}.

\bibitem{Cremmer:1984hj}
E.~Cremmer, C.~Kounnas, A.~Van~Proeyen, J.~P. Derendinger, S.~Ferrara,
  B.~de~Wit et~al., \emph{{Vector Multiplets Coupled to N=2 Supergravity:
  SuperHiggs Effect, Flat Potentials and Geometric Structure}},
  \href{http://dx.doi.org/10.1016/0550-3213(85)90488-2}{\emph{Nucl. Phys.} {\bf
  B250} (1985) 385--426}.

\bibitem{Duff:1995sm}
M.~J. Duff, J.~T. Liu and J.~Rahmfeld, \emph{{Four-dimensional
  string-string-string triality}},
  \href{http://dx.doi.org/10.1016/0550-3213(95)00555-2}{\emph{Nucl. Phys.} {\bf
  B459} (1996) 125--159}, [\href{https://arxiv.org/abs/hep-th/9508094}{{\tt
  hep-th/9508094}}].

\bibitem{Huijse:2011ef}
L.~Huijse, S.~Sachdev and B.~Swingle, \emph{{Hidden Fermi surfaces in
  compressible states of gauge-gravity duality}},
  \href{http://dx.doi.org/10.1103/PhysRevB.85.035121}{\emph{Phys. Rev.} {\bf
  B85} (2012) 035121}, [\href{https://arxiv.org/abs/1112.0573}{{\tt
  1112.0573}}].

\bibitem{Dong:2012se}
X.~Dong, S.~Harrison, S.~Kachru, G.~Torroba and H.~Wang, \emph{{Aspects of
  holography for theories with hyperscaling violation}},
  \href{http://dx.doi.org/10.1007/JHEP06(2012)041}{\emph{JHEP} {\bf 06} (2012)
  041}, [\href{https://arxiv.org/abs/1201.1905}{{\tt 1201.1905}}].

\bibitem{Yuan:2013ts}
F.-F. Yuan and Y.-C. Huang, \emph{{Harrison metrics for the Schwarzschild black
  hole}}, \href{http://dx.doi.org/10.1088/0253-6102/60/5/07}{\emph{Commun.
  Theor. Phys.} {\bf 60} (2013) 551--555},
  [\href{https://arxiv.org/abs/1301.6548}{{\tt 1301.6548}}].

\bibitem{Anninos:2011af}
D.~Anninos, S.~A. Hartnoll and D.~M. Hofman, \emph{{Static Patch Solipsism:
  Conformal Symmetry of the de Sitter Worldline}},
  \href{http://dx.doi.org/10.1088/0264-9381/29/7/075002}{\emph{Class. Quant.
  Grav.} {\bf 29} (2012) 075002}, [\href{https://arxiv.org/abs/1109.4942}{{\tt
  1109.4942}}].

\bibitem{Anantua:2012nj}
R.~J. Anantua, S.~A. Hartnoll, V.~L. Martin and D.~M. Ramirez, \emph{{The Pauli
  exclusion principle at strong coupling: Holographic matter and momentum
  space}}, \href{http://dx.doi.org/10.1007/JHEP03(2013)104}{\emph{JHEP} {\bf
  03} (2013) 104}, [\href{https://arxiv.org/abs/1210.1590}{{\tt 1210.1590}}].

\bibitem{Davison:2014lua}
R.~A. Davison and B.~Goutéraux, \emph{{Momentum dissipation and effective
  theories of coherent and incoherent transport}},
  \href{http://dx.doi.org/10.1007/JHEP01(2015)039}{\emph{JHEP} {\bf 01} (2015)
  039}, [\href{https://arxiv.org/abs/1411.1062}{{\tt 1411.1062}}].

\bibitem{GuicaSkenderisTaylorEtAl2011}
M.~Guica, K.~Skenderis, M.~Taylor and B.~C. van Rees, \emph{{Holography for
  Schrodinger backgrounds}},
  \href{http://dx.doi.org/10.1007/JHEP02(2011)056}{\emph{JHEP} {\bf 02} (2011)
  056}, [\href{https://arxiv.org/abs/1008.1991}{{\tt 1008.1991}}].

\bibitem{CompereGuicaRodriguez2014}
G.~Compère, M.~Guica and M.~J. Rodriguez, \emph{{Two Virasoro symmetries in
  stringy warped AdS$_{3}$}},
  \href{http://dx.doi.org/10.1007/JHEP12(2014)012}{\emph{JHEP} {\bf 12} (2014)
  012}, [\href{https://arxiv.org/abs/1407.7871}{{\tt 1407.7871}}].

\bibitem{Iqbal:2011in}
N.~Iqbal, H.~Liu and M.~Mezei, \emph{{Semi-local quantum liquids}},
  \href{http://dx.doi.org/10.1007/JHEP04(2012)086}{\emph{JHEP} {\bf 04} (2012)
  086}, [\href{https://arxiv.org/abs/1105.4621}{{\tt 1105.4621}}].

\bibitem{Horowitz:1999jd}
G.~T. Horowitz and V.~E. Hubeny, \emph{{Quasinormal modes of AdS black holes
  and the approach to thermal equilibrium}},
  \href{http://dx.doi.org/10.1103/PhysRevD.62.024027}{\emph{Phys. Rev.} {\bf
  D62} (2000) 024027}, [\href{https://arxiv.org/abs/hep-th/9909056}{{\tt
  hep-th/9909056}}].

\bibitem{Berti:2009kk}
E.~Berti, V.~Cardoso and A.~O. Starinets, \emph{{Quasinormal modes of black
  holes and black branes}},
  \href{http://dx.doi.org/10.1088/0264-9381/26/16/163001}{\emph{Class. Quant.
  Grav.} {\bf 26} (2009) 163001}, [\href{https://arxiv.org/abs/0905.2975}{{\tt
  0905.2975}}].

\end{thebibliography}\endgroup

\end{document}